\documentclass[a4paper]{amsart}
\usepackage{latexsym}
\usepackage{amsmath}
\usepackage{amssymb}
\usepackage{amsfonts}

\newcounter{smallarabics}
\newenvironment{arabicenumerate}
{\begin{list}{{\normalfont\textrm{(\arabic{smallarabics})}}}
  {\usecounter{smallarabics}\setlength{\itemindent}{0cm}
   \setlength{\leftmargin}{5ex}\setlength{\labelwidth}{4ex}
   \setlength{\topsep}{0.75\parsep}\setlength{\partopsep}{0ex}
   \setlength{\itemsep}{0ex}}}
{\end{list}}

\newcounter{smallroman}

\newcommand{\ben}{\begin{arabicenumerate}}  
\newcommand{\een}{\end{arabicenumerate}}

\def\init{\setcounter{equation}{0}}

\newtheorem{theoreme}{Theorem }[section]
\newtheorem{proposition}[theoreme]{Proposition}
\newtheorem{lemma}[theoreme]{Lemma}
\newtheorem{definition}[theoreme]{Definition}

\newtheorem{remark}[theoreme]{Remark}
\newtheorem{example}[theoreme]{Example}
\newcommand{\beq}{\begin{equation}}
\newcommand{\eeq}{\end{equation}}

\newcommand{\bex}{\begin{example}}
\newcommand{\eex}{\end{example}}
\def\bel{\begin{lemma}}
\def\eel{\end{lemma}}
\def\bet{\begin{theoreme}}
\def\eet{\end{theoreme}}
\def\bed{\begin{definition}}
\def\eed{\end{definition}}
\def\ber{\begin{remark}}
\def\eer{\end{remark}}

\def\rr{{\mathbb R}}
\def\zz{{\mathbb Z}}
\def\cc{{\mathbb C}}

\def\kk{{\mathbb K}}

\newcommand\SvN{{\rm  SvN}}

\newcommand\en{{\rm en}}
\newcommand\dyn{{\rm dyn}}

\def\Weyl{{\rm Weyl}}
\def\Fin{{\rm FinSym}}
\def\jq{\ii_{\rm ch}}

\def\part{{\rm par}}

\def\Im{{\rm Im}}
\def\Re{{\rm Re}}

\def\cpl{{\rm cpl}}

\def\bar{\overline}

\def\reg{{\rm reg}}

\def\alg{{\rm alg}}

\def\c0inf{C_0^\infty}

\def\s{{\rm s}}
\def\sa{{\rm s/a}}

\def\h{{\rm h}}
\def\ch{{\rm ch}}

\def\cZ{{\mathcal Z}}
\def\cY{{\mathcal Y}}

\def\cW{{\mathcal W}}

\def\fin{{\rm fin}}

\def\CCR{{\rm CCR}}
\def\CAR{{\rm CAR}}

\def\a{{\rm a}}

\def\i{{\rm i}}

\def\Span{{\rm Span}}
\def\Dom{{\rm Dom}}

\def\loplus{\mathop{\oplus}\limits}

\def\sgn{{\rm sgn}}

\def\Ker{{\rm Ker}\,}

\def\gi{{\rm gi}}
\def\fA{{\mathfrak A}}

\def\Sp{{\mathcal Sp}}
\def\12{\frac{1}{2}}
\def\14{\frac{1}{4}}

\def\e{{\rm e}}

\def\d{{\rm d}}
\def\Ran{{\rm Ran}}

\def\bbbone{{\mathchoice {\rm 1\mskip-4mu l} {\rm 1\mskip-4mu l}
{\rm 1\mskip-4.5mu l} {\rm 1\mskip-5mu l}}}
\def\one{\bbbone}
\def\cH{{\mathcal H}}

\def\ii{{\rm j}}

\def\loplus{\mathop{\oplus}\limits}

\def\sgn{{\rm sgn}}

\def\Ker{{\rm Ker}}

\def\cK{{\mathcal K}}

\def\12{\frac{1}{2}}

\def\e{{\rm e}}

\def\d{{\rm d}}
\def\Ran{{\rm Ran}}

\def\cH{{\mathcal H}}

\def\bep{\begin{proposition}}
\def\eep{\end{proposition}}

\def\s{{\rm s}}

\def\CARal{{\rm C\hskip 0.25 em \hbox{\raise 1.72 ex 
\hbox{$\scriptscriptstyle\rm al$}\kern -0.57 em A}R}}

\def\t{{\scriptscriptstyle\#}}
\def\otimesal{\mathop{\hbox{\raise 1.5 ex
  \hbox{$\scriptscriptstyle\rm al$}
\kern -0.92 em \hbox{$\otimes$}}}}
\def\oplusal{\mathop{\hbox{\raise 1.5 ex
  \hbox{$\scriptscriptstyle\rm al$}
\kern -0.92 em \hbox{$\oplus$}}}}
\def\Gammal{\hbox{\raise 1.68 ex 
\hbox{$\scriptscriptstyle\rm al$}\kern -0.50 em $\Gamma$}}
\def\Bal{\hbox{\raise 1.68 ex 
\hbox{$\scriptscriptstyle\rm  al$}\kern -0.50 em $B$}}
\def\CARal{{\rm C\hskip 0.25 em \hbox{\raise 1.72 ex 
\hbox{$\scriptscriptstyle\rm al$}\kern -0.57 em A}R}}
\def\t{{\scriptscriptstyle\#}}

\begin{document}





\title{Positive energy quantization of linear dynamics}

 \author[J. Derezi\'{n}ski, C. G\'erard]{Jan Derezi\'{n}ski\\
Department of Mathematical Methods in Physics,  
\\ Warsaw University,\\ Ho\.{z}a 74, 00-682 Warszawa, Poland,\\
Christian G\'erard\\ D\'epartement de Math\'ematiques \\Universit\'e de
Paris Sud\\ 91405 Orsay Cedex France}

\begin{abstract}The abstract mathematical structure
  behind the positive energy quantization of linear classical
  systems is described.  
It is separated into  3 stages: the description of 
a classical system, the algebraic quantization and the
  Hilbert space quantization.  4 kinds of systems are distinguished:
  neutral bosonic, neutral bosonic, charged bosonic and charged
  fermionic. 

The formalism that is described
follows closely the usual constructions
 employed in quantum physics to
  introduce non-interacting quantum fields.
\end{abstract}

\maketitle

\tableofcontents

\section{Introduction}
In physics, by {\em quantization} one means
various procedures that lead from  {\em  classical} systems to 
{\em quantum} systems. In this paper we describe the basic mathematical
structure of  the {\em positive energy
 quantization} of
 linear  classical systems.  One of
 its  basic requirements is the implementation of the
 quantum dynamics
  by a {\em positive  Hamiltonian}.
In all the cases we consider, the resulting Hilbert space has a
natural structure of a {\em Fock space}, and the dynamics is obtained
by the so-called {\em second quantization} of the dynamics on the
{\em 1-particle space}.

Linear classical systems that we have in mind often have 
an infinite number of degrees of freedom. The most typical examples
are the space  
of solutions of the
{\em Klein-Gordon} and of the {\em Dirac equation},
 possibly on a curved space-time
and in the presence of external potentials. We can also consider 
other systems, not necessarily relativistic, e.g. motivated by the
condensed matter physics at zero temperature.

 The
positive energy quantization, which we
 describe in an abstract fashion
 in this paper, is used
  in quantum field theory  as
the starting point for the construction of 
  {\em
  free } (that means {\em non-interacting})  quantum fields and
  many-body quantum systems.
We will not discuss 
the  quantization of {\em non-linear} dynamics,
 which is usually  more
difficult and often ambiguous.

In quantum physics,
one can distinguish two basic types of particles: {\em bosons} and
{\em fermions}.

Classical theories describing bosons have a natural
{\em  symplectic} structure. After quantization, one obtains
quantum fields  satisfying
{\em canonical commutation relations} (abbreviated 
  CCR). They are usually represented on a {\em bosonic Fock space}.

Classical theories describing fermions
possess a natural {\em euclidean} structure.
 The corresponding quantum fields satisfy
 {\em canonical anticommutation relations} (abbreviated 
  CAR). They are usually represented on a {\em fermionic Fock
   space}. 

Thus classical
bosonic and fermionic systems equipped with a linear dynamics
can be described by a symplectic, resp. euclidean vector space $\cY$.
This space has the interpretation of the {\em dual of the classical phase
space}. (We will not be very pedantic about this point,
 and we will usually call $\cY$
the {\em phase space} as well).  The classical dynamics is described by
 a 1-parameter group $\rr\ni t\mapsto r_t$ of linear symplectic,
resp. orthogonal transformations  on $\cY$.

Both fermionic and bosonic systems appear in two 
varieties: neutral and charged. In the case of charged systems, the
phase space is in addition endowed with an action of the group
$U(1)$. Assuming that it is a representation  of
 {\em charge 1}, it is
natural to encode this symmetry by viewing the phase space as a
{\em complex}
 vector space. In the case of neutral systems, the space $\cY$
is assumed to be {\em real}.

More precisely,  on the classical level charged bosonic systems
are described by
a complex vector space equipped with a {\em nondegenerate
 anti-hermitian} form. We will
call such  spaces {\em charged symplectic}. We will assume that the
dynamics $r_t$ preserves this form.

Charged fermionic systems are described by a {\em unitary} space (complex
space equipped with a positive scalar product) and its dynamics is
 a 1-parameter unitary group.

To sum up, we  distinguish 4 basic formalisms  for quantization:
\ben\item {\bf Neutral bosonic formalism}, applied e.g. to real solutions of the
Klein-Gordon equation;
\item {\bf Neutral fermionic formalism}, applied e.g. to Majorana
  spinors satisfying the 
  Dirac equation; 
\item {\bf Charged bosonic systems}, applied e.g. to complex
 solutions of
  the Klein-Gordon equation; 
\item {\bf Charged fermionic systems}, applied  e.g. to
 Dirac spinors satisfying the
  Dirac equation.
\een

\ber Note that in the 
most common physics applications one uses the neutral
bosonic formalism (e.g. for photons) and the charged fermionic
formalism (e.g. for electrons). Charged bosons are also quite common,
e.g. charged pions or gauge bosons in the standard model.
 On the other hand, until recently,
the neutral fermionic formalism had mostly theoretical interest.
 However, in the modern version of the
standard model  right-handed
massive neutrinos are neutral fermions 
 described by Majorana spinors \cite{Sr}. \eer

One can distinguish 3 stages of quantization:
\ben \item {\bf Classical system.} We consider
 one of the four kinds of the
{\em phase space} $\cY$, together with a 1-parameter group of its
automorphisms, $\rr\ni t\mapsto r_t$,
which we view as a {\em classical dynamics}.
\item {\bf Algebraic quantization.} We choose an  appropriate
  $*$-algebra $\fA$, 
 together with a 1-parameter group of $*$-automorphisms
$\rr\ni t\mapsto \hat r_t$.  The algebra $\fA$ is sometimes called the
 {\em field algebra} of the quantum system. The commutation,
 resp. anticommutation relations satisfied by the appropriate
 distinguished elements of $\fA$ are parallel to relations satisfied by
 elements of the phase space. The 1-parameter group of
 $*$-automorphisms $\hat r_t$ describes
the {\em quantum dynamics in the Heisenberg picture}. The algebra
$\fA$ contains operators that are useful in the theoretical
description of the system. However, we do not assume
that all of its elements are
physically observable, even in principle. Therefore, we 
also  distinguish the {\em algebra of observables}, which
is a certain subalgebra of $\fA$,
invariant with respect to  the dynamics, that
 consists of operators whose measurement is theoretically
 possible. 
\item {\bf Hilbert space quantization.} We represent 
  the algebra $\fA$ on a certain
  Hilbert space $\cH$, so that
 the dynamics is implemented by
  a $1$-parameter unitary group  generated by
  a positive operator, called the {\em Hamiltonian} $H$. Typically,
  this representation is faithful, so that we can write
  $\fA\subset B(\cH)$ and 
\beq \hat r_t(A)=\e^{\i tH}A\e^{-\i tH},\label{imli}\eeq
\een 

The description of quantization that one can find 
in numerous textbooks on quantum field theory 
 is almost always 
presented in a certain concrete context, typically that
of the Klein-Gordon or Dirac equation. In our approach we describe
only the abstract underlying mathematical structure. 
Let us stress, however,
that our presentation, in spite of its abstract mathematical
language,  follows very closely the usual exposition, see
e.g. \cite{We}, and  \cite{Sr};
in particular \cite{Sr} Sec. 22  for complex bosons and \cite{Sr} 
Sec. 49  for 
neutral fermions.

Note that among the three stages of quantization described above, the
most important is the first and the third. The second stage -- the
algebraic quantization -- can be skipped altogether. In the usual
presentation, typical for physics textbooks, it is limited to a formal
level -- one says that ``commuting classical observables'' are
replaced by ``non-commuting quantum observables'' satisfying the
appropriate commutation, resp. anticommutation relations. In our
presentation, we tried to interpret this statement 
in terms of  well defined
$C^*$-algebras. This is quite easy in the case of
fermions. Unfortunately,  in the
case of bosons it leads to certain technical difficulties related to
the unboundedness of bosonic fields. We discuss a number of possible
choices for $C^*$-algebras describing bosonic observables. 
To  some extent, the
algebraic quantization is   merely an exercise of 
 academic
interest. Nevertheless, in some situations it sheds  light on some
conceptual aspects of quantum theory.

One of the confusing 
conceptual points that we believe 
our abstract approach can explain is
the difference between the  phase space and the
1-particle space. Throughout our paper, the former is typically 
denoted by $\cY$
and the latter by $\cZ$. These two spaces are often 
 identified. They have, however, a different physical meaning and are
 equipped with a different algebraic structure.

We also discuss 
 abstract properties
of two commonly used
 discrete symmetries of quantum systems: the time reversal 
and the charge
 reversal. Their properties can be quite confusing. We believe
that the precise
 language of linear algebra is particularly adapted to explain their
 properties.
Note, for instance, that
 the charge reversal is 
 antilinear with respect to the complex structure on the phase space
 and linear with respect to the complex structure on the 1-particle
 space.  On the other hand, the time
reversal is antilinear with respect to both.

Among 
well-known textbooks about rigorous foundations of quantum field
theory one can quote  \cite{BSZ,BR,Em,GJ,Ha,RS2,Si}.
In particular,  \cite{BSZ} contains
a discussion  of some of the aspects of the
 positive energy quantization.
One can argue that our paper collects 
some elements of the knowledge that belongs to the
folklore of theoretical and mathematical physics.
 Nevertheless, a 
systematic and comprehensive discusssion of the topic described in
our paper is to our knowledge difficult to find in the
literature.
We are preparing a monograph \cite{DG2} that will cover it
 in greater detail.

\medskip

{\small {\bf Acknowledgement.}  The research of J.D.
  is  supported in part by the grant N~N201~270135.}
\medskip

\section{Preliminaries}

In this section we introduce a precise
 terminology and notation,
mostly about linear algebra, which we will use in this paper. All of
this section is very elementary. A large part
 of it is standard and well-known. For some concepts we found it
 convenient 
to invent new names. 
The reader may wonder why we need to be so pedantic. 
We will see later on that a surprisingly large variety
of concepts from
basic linear algebra plays an important role in quantization.

\subsection{Vector spaces}
Let $\cY$, $\cW$ be vector spaces over the field $\kk=\rr$ or
$\kk=\cc$. 
$L(\cY,\cW)$ denotes the space of linear maps from $\cY$ to
$\cW$. 

If $\cY$ is a complex space, then $\cY_\rr$ will denote its {\em real form}, 
that is $\cY$ considered as a real space.


\subsection{Symmetric forms}

Let $\cY$ be a  vector space.
Consider a
 bilinear form $\nu$ on $\cY$
\[\cY\times\cY\ni(y_1,y_2)\mapsto y_1\nu y_2\in\kk.\]
We  will say that  $\nu$ is {\em symmetric} if
\[
\ \ \ y_1\nu y_2=y_2\nu y_1,\ \ y_1,y_2\in\cY.
\]

\subsection{Euclidean spaces}

Let $\nu$ be a  symmetric form on a real space $\cY$. It
is called {\em positive} if 
 $y\nu y>0$ for $y\neq0$.
A couple $(\cY,\nu)$, where $\nu$ is a positive form, 
is called a {\em euclidean space}.
If $\cY$ is complete for the
 norm $\|y\|:=\sqrt{y\nu y}$, then it is called
 a {\em real Hilbert space}. 

Let  $(\cY, \nu)$ be a euclidean space and $r\in L(\cY)$. We say
that
\begin{eqnarray*}
r\hbox{ is {\em orthogonal}}&\hbox{ if }&r\hbox{ is bijective and }(ry_1)\nu ry_2=y_1\nu y_2.
\end{eqnarray*}
The set of orthogonal elements in $L(\cY)$ is a group for the operator composition,  
denoted by $O(\cY)$. 

If the form $\nu$ is nondegenerate, but not necessary positive, then
all the  definitions are the same except that 
we add the prefix ``pseudo-'' to the words ``euclidean'', ``isometric''
and ``orthogonal''. 

\subsection{Symplectic spaces}

Let $\cY$ be a   vector space.
We   say that  a bilinear form $\omega$
\[\cY\times\cY\ni(y_1,y_2)\mapsto y_1\omega y_2\in\kk\]
 is {\em anti-symmetric} if
\[
 y_1\omega y_2=-y_2\omega y_1,\ \ y_1,y_2\in\cY.
\]
A nondegenerate antisymmetric bilinear form is  called a {\em
symplectic form}.
The pair $(\cY,\omega)$ is then called  a {\em symplectic space}.

Let $(\cY, \omega)$ be a symplectic space and $r\in L(\cY)$.  We say
that
\begin{eqnarray*}
r\hbox{ is {\em symplectic } if }&&r\hbox{ is bijective and }(r
y_1)\omega r y_2=
y_1\omega y_2,\\
r\hbox{ is {\em anti-symplectic } if }&&r\hbox{ is bijective and }(r
y_1)\omega r y_2=
-y_1\omega y_2
.
\end{eqnarray*}

\subsection{Sesquilinear forms}

Let $\cZ$ be  a complex vector spaces.
Consider a sesquilinear form (anti-linear in the first argument, linear
in the second) 
\[\cZ\times\cZ\ni(z_1,z_2)\mapsto (z_1|\beta z_2)\in\cc.\]
We
 say that \begin{eqnarray*}
\beta\hbox{ is {\em hermitian}}&\hbox{
if}& 
(z_{2}| \beta z_{1})= \bar{(z_1|\beta z_{2})},\: z_1,z_2\in \cZ,
\\
\beta\hbox{ is {\em anti-hermitian}}&\hbox{
if}& 
(z_{2}| \beta z_{1})= -\bar{(z_1|\beta z_{2})},\: z_1,z_2\in \cZ,
\end{eqnarray*}
Clearly, $\beta$ is hermitian iff $\i\beta$ is anti-hermitian.

\subsection{Unitary spaces}

A hermitian form $\beta$  is called {\em positive} if $(z|\beta z)>0$ for
$z\neq 0$.  It is often called
a {\em scalar product}. A pair $\left(\cZ,(\cdot|\beta\cdot)\right)$ is
then sometimes called a unitary space. 
If $\cZ$ is complete for the norm $\|z\|:=\sqrt{(z|\beta z)}$,
 then it is called
 a {\em  Hilbert space}. 

Let $r\in L(\cZ)$.
\begin{eqnarray*}
r\hbox{ is {\em unitary}}&\hbox{ if }&r\hbox{ is  
bijective and }
(r z_1 |\beta rz_2)=
(z_1|\beta z_2).\end{eqnarray*}
The set of unitary 
operators on $\cZ$ is a group  denoted by $U(\cZ)$.

Let
$r\in L(\cZ_{\rr})$ be anti-linear.
 We say
that
\begin{eqnarray*}
r\hbox{ is {\em anti-unitary}}&\hbox{ if }&r\hbox{ is  bijective and }
(r z_1 |\beta rz_2)=
\bar{(z_1|\beta z_2)}.\end{eqnarray*}

\subsection{Charged symplectic spaces}

If $\omega$ is anti-hermitian and non-degenerate,
 then
 $(\cZ,
\omega)$ is called a {\em charged symplectic  space}.

Let $(\cZ, \omega)$ be a charged symplectic
 space  and $r\in L(\cZ)$. We say
that
\begin{eqnarray*}
r\ \ \hbox{ is {\em charged symplectic}}&\hbox{ if }&r\hbox{ is
  bijective and }(r z_1 |\omega r
z_2)= (z_1|\omega z_2),\\
r\ \hbox{ is {\em charged anti-symplectic}}&\hbox{ if }&r\hbox{
   is bijective and }(rz_1| \omega
rz_2)= -(z_1|\omega z_2).
\end{eqnarray*}
The set of charged symplectic 
operators on $\cZ$ is a group for the operator
composition denoted by $ChSp(\cZ)$.

Let
$r\in L(\cZ_{\rr})$ be antilinear.
We adopt the following
 terminology for various kinds of an anti-linear operator
 on a charged symplectic space:
\begin{eqnarray*}
r\ \ \hbox{ is {\em anti-charged symplectic}}&\hbox{ if }&r\hbox{ is
  bijective and }(z_1|\omega z_2)=\bar{(rz_1|\omega rz_2)},\\
r\ \hbox{ is {\em anti-charged anti-symplectic}}&\hbox{ if }&r\hbox{
 is bijective and }(z_1|\omega z_2)=-\bar{(rz_1|\omega rz_2)}.\\
\end{eqnarray*}

\ber The terminology ``charged symplectic space'' is motivated by
applications in quantum field theory: such spaces describe charged
bosons. \eer

\subsection{Complexification of a vector space}
Consider a real space 
$\cY$. Let $\cc\cY$ denote its complexification, that is
$\cc\otimes_\rr\cY$, which is a complex vector space equipped with a
natural conjugation denoted by $\cc\cY\ni w\mapsto \bar w\in \cc\cY$. 

Clearly, every $r\in L(\cY)$ extends uniquely to a linear operator on
$\cc\cY$, which will be denoted $r_\cc$ and to a unique anti-linear
operator on 
$\cc\cY$, which will be denoted $r_{\bar\cc}$.
\subsection{Anti-involutions}

\label{antiin}

Let $\cY$ be  a vector space.
We say that $\ii\in L(\cY)$ is an
{\em anti-involution}, if $\ii^2=-1$.

Let $\cY$ be a real space
  equipped with an anti-involution $\ii\in
L(\cY)$. We can consider it as a complex space, with $\ii$ identified
with the imaginary unit $\i$, and then we will denote it by
$\cY^\cc$. However, we will seldom do so, and in what follows we treat
$\cY$ as a real space.

Note that
 $(\cc\cY)_\rr$ has two distinguished anti-involutions: the usual  $\i$,
and also $\ii_\cc$.
Set
\[
\cZ:=\{y-\i\ii y\ :\
y\in\cY\},\: \ \ \ \bar\cZ:=\{y+\i\ii y\ :\
y\in\cY\}.\]
 $\cZ$ will be called the {\em holomorphic subspace} of
$\cc\cY$,  $\bar\cZ$ will be called the {\em anti-holomorphic subspace}
of $\cc\cY$. 
The
corresponding projections equal
$\one_\cZ:=\12(\one-\i\ii_{\cc})$ and
$\one_{\bar\cZ}:=\12(\one+\i\ii_{\cc})$. Clearly,
 $\one=\one_\cZ+\one_{\bar\cZ}$, and $\cc\cY=\cZ\oplus\bar\cZ$. 
We have $\cZ={\rm Ker}(\ii_{\cc}-\i)$, $\bar{\cZ}= {\rm
Ker}(\ii_{\cc}+\i)$. 

The converse construction is as follows: Let $\cZ$ be a complex
vector 
space. Let $\bar\cZ$ be the space complex conjugate to $\cZ$ (naturally
isomorphic to $\cZ$ as a real space, but with the opposite complex
structure). Set
\[\cY:=\Re(\cZ\oplus\bar\cZ):=\{(z,\bar
z)\in\cZ\oplus\bar\cZ\ :\ z\in\cZ\}.\] 
Clearly $\cY$ is a real vector space  equipped with the
anti-involution
\[
 \ii(z,\bar{z}):= (\i z,  \bar{\i z})= (\i z, -\i \bar{z}).
\]

\subsection{(Pseudo-)K\"ahler spaces}\label{c1.2.6}
Let $(\cdot|\beta\cdot)$ be a hermitian form on a complex space
$\cY$. 
Then on $\cY_{\rr}$
we have a symmetric form $\nu$,
\beq
y_{2}\nu y_{1}:= \Re(y_{2}|\beta y_{1}),
\label{ka1}\eeq
 an anti-symmetric form $\omega$,
\beq
 y_{2}\omega y_{1}:= \Im (y_{2}|\beta y_{1}),
\label{ka2}\eeq
and  an anti-involution $\ii$,
\beq\ii y:=\i y.\label{ka3}\eeq
Note the relationship $y_1\omega\ii y_2=y_1 \nu y_2$.

The name {\em pseudo-K\"{a}hler space} will be
 used for a  space equipped
with a nondegenerate Hermitian form
treated as a real space with the three structures
(\ref{ka1}), (\ref{ka2}) and (\ref{ka3}). Below we give a more precise
definition: 

\begin{definition}
We say that a quadruple $(\cY, \nu, \omega, \ii)$ is a {\em
pseudo-K\"{a}hler space} if\ben
\item $\cY$ is a real vector space,
\item $\nu$ is a nondegenerate symmetric  form,
\item
 $\omega$ is a nondegenerate anti-symmetric  form,
\item
 $\ii$ is an anti-involution,
\item
 $y_1 \omega \ii y_2= y_1\nu y_2$, $y_1,y_2\in\cY$.
\een
If in addition $\nu$ is positive, then we say that
 $(\cY, \nu, \omega, \ii)$ is a {\em K\"ahler space}.
\end{definition}

Two structures out of $ \nu, \omega, \ii$ determine the
third. This motivates the following definitions:
\bed \ben\item
Let $(\cY,\omega)$ be a symplectic space. We say that an
anti-involution $\ii$ is 
 pseudo-K\"ahler if $y_1\omega\ii y_2$ is 
a symmetric form. If in addition it is positive, then we say
that  $\ii$ is K\"ahler.
\item
Let $(\cY,\nu)$ be a euclidean space. We say that an
anti-involution $\ii$ is K\"ahler if
  $y_1\nu\ii y_2$ is  an  anti-symmetric form.
\een\eed

The definitions  (1) and (2) have other equivalent versions, as seen
from the following theorem:
\bet\label{kahler-thm} \ben\item
Let $(\cY, \omega)$ be a symplectic space. 
Then $(\omega,\ii)$ pseudo-K\"ahler iff $\ii\in Sp(\cY)$.
\item
Let $(\cY,\nu)$ be a euclidean space.
Then $(\nu,\ii)$ is K\"ahler iff $\ii\in O(\cY)$.
\een
\eet

\subsection{$U(1)$ symmetries of charge 1}

Let $\cY$ be a real space. Let $U(1)$ be the group $\rr/2\pi\zz$.
Let $U(1)\ni \theta\mapsto u_\theta\in L(\cY)$
be a representation. 
\bed We say that it is a {\em representation of charge 1}
if there exists an anti-involution $\jq$ such that
\[u_\theta=\cos\theta\one+\sin\theta\jq.\]
\eed

\bep Let $(u_\theta)_{\theta\in U(1)}$ be a representation of charge
1.
\ben\item
Assume that  
$\cY$ is a symplectic space. Then $u_\theta$ is
symplectic for $\theta\in U(1)$ iff $\jq$ is pseudo-K\"ahler.
\item
Assume that  
$\cY$ is a euclidean space. Then $u_\theta$ is
orthogonal for $\theta\in U(1)$ iff $\jq$ is K\"ahler.
\een\eep

\subsection{Operators on Hilbert spaces}

If  $\cH$, $\cK$ are Hilbert spaces, then
$B(\cH,\cK)$, $U(\cH,\cK)$, resp. $Cl(\cH,\cK)$
 denotes the space of bounded, unitary, resp. closed
 operators from $\cH$ to
$\cK$. 
 We set $B(\cH):=B(\cH,\cH)$,  $U(\cH):=U(\cH,\cH)$,  $Cl(\cH):=Cl(\cH,\cH)$.

 $B_\h(\cH)$,    resp. $Cl_\h(\cH)$
denotes the set of bounded self-adjoint, resp. 
closed self-adjoint operators
on $\cH$. 

If $(\cZ_i)_{i\in I}$ is a family of Hilbert spaces, then 
$\loplus_{i\in I}\cZ_i$ will always denote the direct sum in the sense
of Hilbert spaces.

If $\cZ,\cW$ are Hilbert spaces, then $\cZ\otimes\cW$ will always
denote the tensor product of $\cZ$ and $\cW$ in the sense of Hilbert
spaces.

\section{Canonical commutation
and anticommutation relations}\init\label{c5}

In this section first we will 
discuss the concept of a representation of canonical
commutation relations, abbreviated a CCR representation. Then we will
introduce the notion of a
representation of canonical
anticommutation relations, abbreviated a CAR representation. Both
concepts come in two varieties:  neutral and charged.

CCR and CAR representations
 have a long history, going back to e.g.
\cite{Di1,JW}.  They were for quite some time an important subject of
research in mathematical physics, let us mension
\cite{Ar1,Ar2,Ar3,ArShi,CMR,Sla}. 
For a textbook reference on CCR and CAR representations let us quote
 \cite{BR},
see also \cite{De}.

Throughout this section,  $\cH$ will denote  a Hilbert space.

\subsection{CCR representations}\label{c5.1}
 
Let $(\cY,\omega)$ be a symplectic space. Let us first try
 to define
 the concept of a CCR representation  in a  naive way. We would like
 to have
 a linear map
\beq \cY\ni y\mapsto \phi^\pi(y)\in Cl_\h(\cH)\eeq
satisfying 
 \beq
[\phi^\pi(y_1),\phi^\pi(y_2)]=\i y_1\omega y_2\one.\label{hei0}\eeq
We will call 
 (\ref{hei0}) the {\em canonical
 commutation
  relation in the Heisenberg form}. 
Unfortunately, this relation is
 problematic from the rigorous point of view,
 because one needs to
supply it with the precise meaning of the commutator of unbounded
operators on the left hand side.

Weyl proposed to replace  (\ref{hei0}) with another
 relation involving
the operators $\e^{\i\phi^\pi(y)}$. These operators are
bounded, and therefore one does not need to discuss domain 
questions. In our
definition of CCR representations we will use the canonical
commutation relations in the so-called
Weyl form.
Under  additional
regularity assumptions they  imply the CCR in the Heisenberg form.

\bed  A
 {\em representation of the canonical commutation relations} or 
a {\em CCR
 representation over $(\cY, \omega)$ in $\cH$}
is a map
\beq
\cY\ni y\mapsto W^\pi(y)\in U(\cH)
\label{iui}\eeq satisfying
\beq\begin{array}{l}W^\pi(y_1)W^\pi(y_2)=\e^{-\frac{\i}{2} y_1\omega y_2}
W^\pi(y_1+y_2).
\end{array}\label{weyl1}\eeq
$W^\pi(y)$ is then called the {\em Weyl operator} corresponding to
$y\in\cY$. 
\label{defccr}\eed


\subsection{Regular CCR representations}

\bed  A CCR representation  (\ref{iui}) is called
{\em regular}  if
\beq\rr\ni t\mapsto W^\pi(ty)\in U(\cH)\ \ \ \hbox{ is strongly continuous for
any $y\in \cY$}.\label{gro}\eeq \eed

Clearly, $\rr\ni t\mapsto W^\pi(ty)$
 is a strongly continuous 1-parameter unitary group.
By the Stone theorem, for any $y\in \cY$, we can define its
self-adjoint generator
\[\phi^\pi(y):=-\i\frac{\d}{\d t}W^\pi(ty)\Big|_{t=0}.\]
In other words, $\e^{\i\phi^\pi(y)}=W^\pi(y)$.

\bed
$\phi^\pi(y)$ will be called the  {\em (bosonic) field operator}
corresponding to $y\in\cY$. \eed



Let $w\in\cc\cY$. We can write
$w=y_1+\i y_2$ for $y_1,y_2\in\cY$. We set
\[\phi^\pi(w):=\phi^\pi(y_1)+\i\phi^\pi(y_2).\]
with
$\Dom \phi^{\pi}(w):=\Dom \phi^{\pi}(y_{1})\cap \Dom
\phi^{\pi}(y_{2})$. 
\bed
$\phi^\pi(w)$ will be called the  {\em complex field operator}
corresponding to $w\in\cc\cY$. \eed

It is easy to show the following proposition:

\bep\label{thm-fields} Let $y,y_1,y_2\in\cY$. 
\ben \item $\phi^\pi(y)$ are closed.
\item $\phi^\pi(ty)=t\phi^\pi(y)$, $t\in\rr$.
\item On $\Dom \phi^{\pi}(y_{1})\cap \Dom
\phi^{\pi}(y_{2})$
we have $\phi^\pi(y_1+\i y_2)=\phi^\pi(y_1)+\i 
\phi^\pi(y_2)$.
\item $[\phi^\pi(y_1),\phi^\pi(y_2)]=\i y_1\omega y_2\one$
 holds as a quadratic form on $\Dom \phi^{\pi}(y_{1})\cap \Dom
\phi^{\pi}(y_{2})$.
\een\eep

\subsection{Charged CCR representations}
\label{c5.2.3c}
CCR representations, as defined in  Def. \ref{defccr},
are used mainly to describe neutral bosons.
Therefore, sometimes we will call them {\em neutral CCR
  representations}. 
In the context of charged bosons one uses another formalism
described in the following definition.

Let $\left(\cY,(\cdot|\omega\cdot)\right)$
 be a charged symplectic space.
 Let us first try
 to define a charged CCR representation in a naive way. We would like
 to have a linear map $\cY\ni y\mapsto\psi^\pi(y)\in Cl(\cH)$
 satisfying 
\begin{eqnarray}\nonumber
[\psi^{\pi*}(y_1),\psi^{\pi*}(y_2)]&=&
[\psi^{\pi}(y_1),\psi^{\pi}(y_2)]=0,\\{}
[\psi^{\pi}(y_1), \psi^{\pi*}(y_2)]&=&\i(y_1|\omega y_2)\one,\ \ \ \  
\ y_1,y_2\in\cY.\label{naive}
\end{eqnarray}
Again, the above definition is problematic.
A possible rigorous definition is given below:

\bed  We say
that a  map
\begin{eqnarray}
\cY\ni y&\mapsto&\psi^\pi(y)\in Cl(\cH)
\label{char}\end{eqnarray}
is a charged CCR  representation iff
there exists a map
\begin{eqnarray*}
\cY\ni y&\mapsto&\phi^\pi(y)\in Cl_\h(\cH)
\end{eqnarray*}
such that
\[\e^{\i\phi^\pi(y_1)}\e^{\i\phi^\pi(y_1)}=
\e^{-\frac{1}{2}\Re(y_1|\omega
  y_2)}\e^{\i\phi^\pi(y_1+y_2)},\ y_1,y_2\in\cY, \]
 $\phi^\pi(ty)=t\phi^\pi(y)$, $t\in\rr$, 
 $\Dom\,\psi^\pi(y)=
\Dom\,\phi^\pi(y)\cap \Dom\,\phi^\pi(\i y)$ and
\[\psi^{\pi}(y)=\frac{1}{\sqrt2}\left(\phi^\pi(y)+\i\phi^\pi(\i y)\right),
\ \ y\in\cY.
\] 
\eed

Note that a charged CCR  representation satisfies the  conditions
of the ``naive definition'':
\bep  Consider a charged CCR representation.
Let $y,y_1,y_2\in\cY$. 
\ben \item $\psi^\pi(\lambda y)=\lambda\psi^\pi(y)$,
$\lambda\in\cc$; 
\item On $\Dom\psi^\pi(y_1)\cap\Dom\psi^\pi(y_2)$ we have
$\psi^\pi(y_1+y_2)=\psi^\pi(y_1)+\psi^\pi(y_2)$;
\item In the sense of quadratic forms, we have the identities
\begin{eqnarray*}[\psi^{\pi*}(y_1),\psi^{\pi*}(y_2)]&=&
[\psi^{\pi}(y_1),\psi^{\pi}(y_2)]=0,\\{}
[\psi^{\pi}(y_1), \psi^{\pi*}(y_2)]&=&\i(y_1|\omega y_2)\one,\ \ \ \  
\ y_1,y_2\in\cY.
\end{eqnarray*}
\een\eep
 
Note that to any charged CCR representation (\ref{char}) we can
associate a regular neutral CCR representation over $\cY$ equipped
with $\Re(\cdot|\omega\cdot)$
\begin{eqnarray*}
\cY\ni y&\mapsto&\e^{\i\phi^\pi(y)}\in U(\cH),
\end{eqnarray*}
as well as a $U(1)$ symmetry of charge $1$
\[U(1)\ni\theta\mapsto \e^{\i\theta}\one\in Sp(\cY).\]

Conversely, 
charged CCR representations arise when we have a (neutral)
CCR representation and the underlying symplectic space
is equipped with a charge 1 symmetry. Let us make this
precise.  Suppose that $(\cY,\omega)$ is a symplectic space and
\[\cY\ni y\mapsto \e^{\i\phi(y)}\in U(\cH)\]
a neutral CCR representation. Suppose that
 $\jq$ is
a pseudo-K\"ahler anti-involution, so that $U(1)\ni\theta\mapsto
u_\theta=\cos\theta\one+\sin\theta \jq\in Sp(\cY)$ is a charge 1 
symmetry.
Following 
 Subsection \ref{antiin},  we
 introduce the holomorphic subspace for $\jq$, that is
 \[\cZ_\ch:=\{y-\i\jq y\ :\ y\in\cY\}\subset\cc\cY.\]
We have a natural identification of the space $\cZ_\ch$ with $\cY$:
\[\cY\ni y\mapsto z= \frac{1}{\sqrt2}(\one-\i\jq)y.\]
We use this identification to define charged fields
parametrized by
$\cY$: 
\[\psi^{\pi*}(y):=
\phi^{\pi}\left(z\right),\ \ \ 
\psi^\pi(y):=
\phi^{\pi}\left(\bar z\right).\]
Thus we obtain a charged CCR representation over $\cY^\cc$ with the
complex structure given by $\jq$ and the anti-hermitian form
\[(y_1|\omega y_2):=y_1\omega y_2-\i y_1\omega\jq y_2.\]

\subsection{CAR representations}\label{c12.1.1}

Let $(\cY, \nu)$ be a 
euclidean space, that is a real vector space  $\cY$
equipped with a positive
symmetric form $\nu$.

In this subsection we   introduce the concept of a representation of
canonical anticommutation relations.
The definition that we use is very similar to the well-known
 definiton of
a representation of Clifford relations.
 In the case of CAR representations we assume in addition that the
operators satisfying the Clifford relations  act on a Hilbert space
and are self-adjoint.

CAR representations appear in quantum physics in at least  two
contexts. First, they describe  many body fermionic systems. 
 Second, they describe spinors, that
is, representations of  Spin groups. In most applications the second
meaning is restricted to the finite dimensional case.

Recall that 
$[A,B]_+:=AB+BA$ is the anticommutator of $A$ and $B$. 
\bed  A {\em representation 
of the canonical anticommutation relations} or a {\em  CAR representation
 over $\cY$ in $\cH$}  is a linear
map
\beq\cY\ni y\mapsto\phi^\pi( y)\in B_\h(\cH)\label{car1a}\eeq
 satisfying
\beq[\phi^\pi(y_1),\phi^\pi(y_2)]_+= 2 y_1\nu
y_2\one,\ \ \ y_1,y_2\in\cY 
.\label{car2a}\eeq 
The operators $\phi^{\pi}(y)$ are called {\em (fermionic)  field
operators}.

\label{car2a1}
\eed

\ber
 Unfortunately, the analogy between the
CAR (\ref{car2a}) and the CCR  (\ref{hei0}) is somewhat spoiled by the
number $2$ on the right hand side of (\ref{car2a}).
The reason for this  convention
is to have
the identity  $\phi^\pi(y)^2=y\nu y\one$.\eer

In what follows we assume that we are given a CAR representation 
(\ref{car1a}).
The operators $\phi^{\pi}(y)$ are called (fermionic) {\em field
operators}.
By complex linearity we can extend the definition of  field operators to
$w=y_1+\i y_2\in\cc\cY$, where $ y_1,y_2\in\cY$:
\[
\phi^\pi(w):=\phi^\pi(y_1)+\i\phi^\pi( y_2).
\]
\bed The operators $\phi^{\pi}(w)$ for $w\in \cc\cY$ are called {\em
complex field operators}. \eed

\subsection{Charged CAR representations}

The concept of CAR relations, as defined in  Def. \ref{car2a},
 is used mainly to describe neutral fermions.
Therefeore, sometimes we will call them {\em neutral CAR
  representations}. 
In the context of charged fermions one uses another formalism
described in the following definition.

 Suppose that $\left(\cY,(\cdot|\cdot)\right)$ is a unitary space.

\bed  We say
that a linear map\begin{eqnarray*}
\cY\ni y&\mapsto&\psi^{\pi}(y)\in B(\cH)
\end{eqnarray*}
is a charged CAR  representation iff
\begin{eqnarray*}
[\psi^{\pi*}(y_1),\psi^{\pi*}(y_2)]_{+}&=&
[\psi^{\pi}(y_1),\psi^{\pi}(y_2)]_{+}=0,\\{}
[\psi^{\pi}(y_1), \psi^{\pi*}(y_2)]_{+}&=&(y_1|y_2)\one,\ \ \ \  
\ y_1,y_2\in\cY.
\end{eqnarray*}
\eed

Suppose that
 $y\mapsto\psi^\pi(y)$ is a charged CAR
representation. Set
\beq\phi^\pi(y):=\psi^\pi(y)+\psi^{\pi*}(y),\eeq
\beq y_1\nu y_2:=\Re(y_1|y_2).\eeq
Then  $\cY\ni y \mapsto \phi^\pi(y)\in B_\h(\cH)$
is a neutral CAR representation over the euclidean space $(\cY,\nu)$.
In addition, $\cY$ is equipped with a 
charge 1  symmetry $U(1)\ni\theta\mapsto\e^{\i\theta}\one\in
  O(\cY_\rr)$. 

Conversely, charged CAR representations
 arise when we have a (neutral)
CAR representation and the underlying euclidean space
is equipped with a  $U(1)$ symmetry of charge 1.
 Let us make this
precise.  Suppose that $(\cY,\nu)$ is a euclidean space and
\[\cY\ni y\mapsto \phi^\pi(y)\in B_\h(\cH)\]
is a neutral CAR representation. Suppose that 
 $\jq$ is a K\"ahler anti-involution, so that
$U(1)\ni\theta\mapsto
u_\theta=\cos\theta\one+\sin\theta\jq\in O(\cY)$ is a charge 1
symmetry. 
Following  Subsect, \ref{antiin},  we
 introduce the holomorphic subspace for $\jq$, that is
 \[\cZ_\ch:=\{y-\i\jq y\ :\ y\in\cY\}\subset\cc\cY.\]
We have a natural identification of the space $\cZ_\ch$ with $\cY$:
\[\cY\ni y\mapsto z= \frac{1}{2}(\one-\i\jq)y.\]
We use this identification to define
charged fields as
\[\psi^{\pi*}(y):=
\phi^{\pi}\left(z\right),\ \ \ 
\psi^\pi(y):=
\phi^{\pi}\left(\bar z\right).\]
Thus we obtain a charged CAR representation over $\cY^\cc$ with the
complex structure given by $\jq$ and the scalar product
\beq
(y_1|y_2):=y_1\nu y_2-\i y_1\nu\jq
y_2,\ \ y_1,y_2\in\cY.\label{sese}\eeq

\section{Fock spaces}

In this section we fix our terminology related to bosonic and
fermionic Fock spaces. In particular, we introduce the so-called Fock
CCR and CAR representations.
 Unfortunately, no
uniform notation concerning this material 
seems to exist in the literature. 

Let us quote, for
example,  the
following works which discuss constructions related to Fock spaces:
\cite{BR,De,DG1,GJ,RS2}

\subsection{Tensor algebra}\label{c3.2.1}
Let $\cZ$ be a Hilbert space space.
Let $\otimes^n\cZ$  denote the $n$th
 tensor power of $\cZ$. 
We set $\otimes^0\cZ:=\cc$.
The {\em complete tensor algebra} over
$\cZ$ is defined as
\[
 \otimes
\cZ:=\loplus_{n=0}^\infty \otimes^n\cZ,\]
It is also sometimes  called the {\em full Fock space}.

The element $1\in\otimes^0\cZ$ is  called
 the {\em vacuum} and will be denoted by $\Omega$.

\subsection{Operators $\d\Gamma$ and $\Gamma$
on the tensor algebra}\label{c3.2.2}

Let
 $\cZ,\cZ_{1}, \cZ_{2}$ be Hilbert spaces.
Suppose that  $p\in B(\cZ_1,\cZ_2)$ is a contraction. We define
\begin{eqnarray*}
\Gamma^n(p)&:=&p^{\otimes n}\ \ \ \in \ \ \ B(\otimes^n\cZ_1,\otimes^n\cZ_2),\\
\Gamma(p)&:=&\loplus_{n=0}^\infty \Gamma^n(p) \in 
B(\otimes\cZ_1,\otimes\cZ_2).\end{eqnarray*}

Likewise, if $h\in Cl(\cZ)$ then we define 
\begin{eqnarray*}
\d\Gamma^n(h)&:=&
\sum_{j=1}^n\one_\cZ^{\otimes j-1}\otimes h\otimes
\one_\cZ^{\otimes(n-j)}\in Cl(\otimes^n\cZ),\\
\d\Gamma (h)&:=&\loplus_{n=0}^\infty\d \Gamma^n(h)\ \ \in\ \ 
Cl(\otimes\cZ).\end{eqnarray*}

The {\em number operator} and the 
 {\em parity operator} are defined respectively as
\begin{eqnarray}
N&:=&\d\Gamma(\one),\label{number}\\
I&:=&(-1)^N=\Gamma(-\one).\label{c1para0}\end{eqnarray}

\bep
 Let $h,h_1,h_2\in B(\cZ)$,
$p_1\in B(\cZ,\cZ_1)$, $p_2\in B(\cZ_1,\cZ_2)$, $\|p_1\|,\|p_2\|\leq1$. We then have
\begin{eqnarray*}
\Gamma(\e^{ h})&=&\e^{\d\Gamma(h)},\\
\Gamma(p_2)\Gamma(p_1)&=&\Gamma(p_2 p_1),\\{}
[\d\Gamma(h_1),\d\Gamma(h_2)]&=&\d\Gamma([h_1,h_2]).\end{eqnarray*}
\eep

\subsection{Bosonic and fermionic Fock spaces}
\label{c3.2.3} 
 
Let $\cZ$ be a Hilbert space. 
If $\sigma\in S_n$, then there exists a unique
\[\Theta(\sigma)\in U(\otimes^n\cZ_1)\]
such that
\[\Theta(\sigma)y_1\otimes\cdots\otimes y_n
=y_{\sigma^{-1}( 1)}\otimes\cdots\otimes
y_{\sigma^{-1}( n)}.\]
Clearly,
\[S_n\ni\sigma\mapsto
 \Theta(\sigma)\in U(\otimes^n\cZ)\]
is
 a representation of the permutation group.
 We define the following operators on $\otimes^n\cZ$:
 \begin{eqnarray*}
\Theta_\s^n&:=&\frac{1}{n!}\sum_{\sigma\in S_{n}}\Theta(\sigma)
,\\
\Theta_\a^n&:=&\frac{1}{n!}\sum_{\sigma\in S_{n}}\sgn\sigma\Theta(\sigma).\end{eqnarray*}
It is easy to check that
 $\Theta_\s^n$ and $\Theta_\a^n $ are orthogonal 
projections.

We will write $\sa$ as a subscript which can mean either $\s$ or
$\a$. 
 We set
\begin{eqnarray*}
\Gamma_\sa^n(\cZ)&:=&\Theta_\sa^n\otimes^n\cZ,\\
\Gamma_\sa(\cZ)&:=&\oplus_{n=0}^\infty\Gamma_\sa^n(\cZ)
=\Theta_\sa
\otimesal\cZ.\end{eqnarray*}
$\Gamma_\sa(\cZ)$ are called the 
 {\em bo\-so\-nic, resp. fermionic   Fock space} \cite{Fo}.

Occasionally, we will need the {\em finite particle
 bo\-so\-nic, resp. fermionic   Fock spaces}, denoted
$\Gamma_\sa^\fin(\cZ)$,
 which are the subspaces of $\Gamma_\sa(\cZ)$ consisting of
finite sums of $n$-particle vectors.

\subsection{$\d\Gamma$ and $\Gamma$ operators on Fock spaces}\label{c3.2.5}

If $p\in B(\cZ,\cW)$ is a contraction,
 then $\Gamma^n(p)$ maps $\Gamma_\sa^n(\cZ)$
 into $\Gamma_\sa^n(\cW)$. Hence $\Gamma(p)$ maps
$\Gamma_\sa(\cZ)$
 into  $\Gamma_\sa(\cW)$. We will use the same symbols 
 $\Gamma^n(p)$ and  $\Gamma(p)$ to denote the corresponding 
restricted operators. $\Gamma(p)$ is sometimes called the {\em second
  quantization of $p$}.

If $h\in Cl(\cZ)$, then $\d\Gamma^n(h)$ maps $\Gamma_\sa^n(\cZ)$
 into itself. Hence, $\d\Gamma(h)$ maps $\Gamma_\sa(\cZ)$
 into itself. We will use the same symbols 
 $\d\Gamma^n(h)$ and  $\d\Gamma(h)$ to denote the corresponding 
restricted operators. Perhaps, the correct name of $\d\Gamma(h)$
should be the {\em infinitesimal second quantization of $h$}.

\subsection{Creation and annihilation operators}

Let $\cZ$ be a Hilbert space and $z\in\cZ$. 
We consider the bosonic or fermionic Fock space $\Gamma_\sa(\cZ)$.

Let $z\in\cZ$. 
We will now 
define two operators with the domain $\Gamma_\sa^\fin(\cZ)$.

The {\em creation operator} of $z$ is defined as
\[
c(z)\Psi:=\sqrt{n+1}\Theta_\sa^{n+1}z\otimes\Psi,\ \ \Psi\in\Gamma^n_\sa(\cZ).
\]
The
 {\em annihilation operator} of $z$, satisfies
\[
a(z)\Psi:=\sqrt{n}(z|{\otimes} \one\ \Psi,\ \ \Psi\in\Gamma^n_\sa(\cZ).
\]

\bep\ben\item
In the bosonic case, the operators $ c(z)$ and $a(z)$ are densely defined and
closable. We denote  their closures by the same symbols. They satisfy
 $a( z)^*= c(z)$.
 Therefore we will write $ a^*(z)$ instead of $c(z)$. 
\begin{eqnarray*}
[a^*(z_1),a^*(z_2)]&=&0,\ \ \ 
[a( z_1),a( z_2)]\ =\ 0,\\{}
[a( z_1),a^*(z_2)]&=&(z_1|z_2)\one.\end{eqnarray*}
\item
In the fermionic
 case, the operators $ c(z)$ and $a(z)$ are densely defined and
 bounded. 
We denote by the same symbols their closures. They satisfy
 $a( z)^*=c(z)$.
 Therefore we will write $ a^*(z)$ instead of
$c(z)$.
We have
\begin{eqnarray*}
[a^*(z_1),a^*(z_2)]&=&0,\ \ \ 
[a( z_1),a( z_2)]\ =\ 0,\\{}
[a( z_1),a^*(z_2)]&=&(z_1|z_2)\one.\end{eqnarray*}
\een\eep

\subsection{Fock CCR representation}

For a Hilbert space $\cZ$, 
 we introduce
the space \[\cY=\Re(\cZ\oplus\bar\cZ):=
\{(z,\bar z)\ :\ z\in\cZ\},\]
which will  serve as the {\em phase space} of our system.
It will be equipped with a symplectic form $\omega$ and
a K\"ahler  anti-involution $\ii$:
 \begin{eqnarray}\label{symplec}
(z,\bar z)\omega (w,\bar w)&:=&2\Im(z|w),
\\
\label{sca1}
\ii(z,
\bar z)&:=& (\i z, \bar{\i z}).
\end{eqnarray}


\bep
\beq
\cY\ni y\mapsto W(y)=\e^{\i a^*(z)+\i a(z)}\in
U(\Gamma_\s(\cZ)),\ 
y=(z,\bar z).\label{focko}\eeq is a regular CCR representation
 over $(\cY,\omega)$ on 
$\Gamma_\s
(\cZ)$. \eep
\bed
We call (\ref{focko}) the {\em Fock CCR representation}. 
\eed

\subsection{Fock CAR representation}

Let $\cZ$, $\cY$ and $\ii$ remain as in previous subsection.
We equip $\cY$
with the structure of a euclidean space with the scalar product
$\nu$: 
\begin{eqnarray*}
(z,\bar z)\nu (w,\bar w)&:=&\Re(z|w),\\
\label{sca10}\end{eqnarray*}
Clearly, $\ii$ is a K\"ahler anti-involution  for $\nu$.

\bep \beq
\cY\ni y\mapsto \phi(y)= a^*(z)+ a(z)\in
B_\h(\Gamma_\s(\cZ)),\ 
y=(z,\bar z).\label{focko5}\eeq
is a CAR representation over $(\cY,\nu)$ on 
$\Gamma_\a
(\cZ)$. \eep 
\bed We call (\ref{focko5}) the {\em Fock CAR 
representation}. \eed

\section{CCR and CAR algebras}

In some approaches to quantum physics a considerable importance is
attached to the choice of a
$*$-algebra, usually a $C^*$- or $W^*$-algebra, which is supposed to
describe observables of a system \cite{BR,Em,Ha}. By choosing a state (or a family of
states) and making the corresponding GNS construction, we obtain a
representation of this $*$-algebra in a Hilbert space.
This philosophy allows us to study
quantum systems in a representation independent fashion.

Many authors try to apply this to bosonic and fermionic systems.  
This is especially natural in the case of fermionic systems,
where there
 exists an obvious choice of a $C^*$-algebra
describing the CAR  over a given Euclidean  space.

In the bosonic case the situation is 
more problematic. In particular,  for a given symplectic space
several  natural choices of CCR
algebras are possible. We will describe some of them.

\subsection{Weyl CCR algebras}

\label{s.weylo}

Let
 $(\cY,\omega)$ be a symplectic space, not necessarily of finite
dimension. 
In this section we introduce the notion of the Weyl CCR $C^*$-algebra
over $\cY$. It is
the $C^*$-algebra generated by elements satisfying the Weyl CCR
relations over $\cY$. 
Many mathematical physicists use them in their description
of bosonic systems. 

Let us start with the definition of the {\em algebraic Weyl CCR
  algebra over $\cY$} .

\bed  
$\CCR_\alg^\Weyl(\cY)$  is defined as
 the $*$-algebra with a basis
given by elements $W(y)$, $y\in\cY$, satisfying the relations
\begin{eqnarray*}W(y_1)W(y_2)&=&\e^{-\frac{\i}{2} y_1\omega y_2}
W(y_1+y_2),\ \ y_1,y_2\in\cY;\\
W(y)^*&=&W(-y),\ \ \ \ \ \ \ \ \ \ \  \ y\in\cY.
\end{eqnarray*} \eed

The following theorem comes from \cite{Sla}, see also \cite{BR}:
\bet There exist faithful $*$-representations of
 $\CCR_\alg^\Weyl(\cY)$.
Let $A\in \CCR_\alg^\Weyl(\cY)$ and let 
$\pi$ be such a representation.
Then $\|\pi(A)\|$ does not depend on $\pi$.
\eet

Thus  $\CCR_\alg^\Weyl(\cY)$ possesses a unique $C^*$-norm.

\bed 
The  {\em Weyl CCR $C^*$-algebra} is defined as
\[\CCR^\Weyl(\cY):=\left(\CCR_\alg^\Weyl(\cY)\right)^\cpl.\]
\eed
where ${\scriptstyle \cpl}$ denotes the completion. Clearly, 
$\CCR^\Weyl(\cY)$ is a $C^*$-algebra.

The following isomorphisms are sometimes called {\em Bogoliubov
  automorphisms}.

\bep 
Let $r\in
Sp(\cY)$. Then there exists a
unique  $*$-automorphism
$\hat r:\CCR^\Weyl(\cY)\to\CCR^\Weyl(\cY)$ such that $\hat r(W(y))=W(ry)$,
$y\in \cY$.
 \label{pres3}\eep

The following proposition explains the relationship between CCR
representations and CCR algebras.
\bep Let
  $\cY\ni y\mapsto W^\pi(y)\in U(\cH)$ be a 
CCR representation. Then there exists a unique 
 $*$-representation 
$\pi:\CCR^\Weyl(\cY)\to B(\cH)$ such that $\pi(W(y))=W^\pi(y)$.
Moreover, $\pi$ is isometric. \eep

\subsection{Stone -- von Neumann
CCR algebras}

One can argue that the Weyl CCR algebra is somewhat artificial. In
particular, it is a noncommutative analogue of the space of almost
periodic functions, which is quite a pathological object.
One can therefore try to look for alternatives for Weyl CCR algebras.
In this and the next subsection we discuss some alternative algebras
describing CCR.

In this subsection we always assume that $(\cY,\omega)$
 is a finite dimensional symplectic space. 
Let us recall the Stone-von Neumann Theorem about the uniqueness of
regular CCR relations (see e.g. \cite{BR}).

\bet Suppose that, for $i=1,2$, $\cH_i$ are  Hilbert spaces and
\[\cY\ni y\mapsto W_i(y)\in U(\cH_i)\]
are regular CCR representations such that
$\Span\{W_i(y)\ :\ y\in\cY\}$ is weakly dense
in $ B(\cH_i)$.
 Then there exists 
 $U\in U(\cH_1,\cH_2)$,
 unique up to a  phase factor,
 such that $W_2(y)=UW_1(y) U^*$, $y\in\cY$.
 \label{alga2}\eet

Theorem \ref{alga2} suggests the following definition:

\bed
A {\em Stone-von Neumann CCR algebra over $\cY$}
is defined as the von Neumann  algebra $B(\cH)$ for a certain Hilbert
space $\cH$
with distinguished unitary elements
$W(y)$, $y\in\cY$, 
such that $\cY\ni y\mapsto W(y)$ is a regular CCR representation and
$\Span\{W(y)\ :\ y\in\cY\}$ is weakly dense in $B(\cH)$.
 It is denoted $\CCR^\SvN(\cY)$.
\label{stvnal}\eed

By Theorem \ref{alga2},  $\CCR^\SvN(\cY)$ is defined uniquely up to a
unitary equivalence.
Clearly,
 $\CCR^\SvN(\cY)$ is not very interesting  as a von Neumann-algebra 
-- it is isomorphic to the usual type I factor. What is interesting
is the category of Bogoliubov automorphisms between these
algebras, described in the following proposition:

\bep
\ben \item
Let $r\in
Sp(\cY)$. Then there exists a
unique spatially implementable $*$-automorphism
$\hat r$ of  $\CCR^\SvN(\cY)$ such that $\hat r(W(y))=W(ry)$,
$y\in \cY$.
\item Let $\cY_1$ be a symplectic subspace of $\cY$. 
Then there is a unique embedding of
$\CCR^\SvN(\cY_1)$
 in  $\CCR^\SvN(\cY)$, 
 such that, for $y\in\cY_1$,
$W(y)$ in the sense of  $\CCR^\SvN(\cY_1)$ coincide with $W(y)$
 in the sense of
  $\CCR^\SvN(\cY)$.
\een
 \label{pres1}\eep

\subsection{Regular CCR algebras}
\label{s.reg}

Let
 $(\cY,\omega)$ be a symplectic space of  arbitrary
dimension. $\Fin(\cY)$ will denote the set of finite dimensional
symplectic subspaces of $\cY$. 

In this subsection we introduce the notion of the regular CCR
algebra  over $\cY$. In the literature, it is rarely
used. Weyl CCR algebras are more common.
 Nevertheless, it is a  natural construction, and we will find it
 useful, especially to describe charged bosonic systems. Its use was
 advocated by I.Segal.

Let $\cY_1,\cY_2\in\Fin(\cY)$ and $\cY_1\subset\cY_2$. 
We can define their Stone-von Neumann
 CCR algebras, as in Definition
\ref{stvnal}. By Proposition \ref{pres1}, we have a natural embedding
\[\CCR^\SvN(\cY_1)\subset\CCR^\SvN(\cY_2).\]

We can define the algebraic regular CCR $*$-algebra as the inductive
limit of  Stone-von Neumann
 CCR algebras:

\bed We set
\beq\CCR_\alg^\reg(\cY):=\bigcup_{\cY_1\in\Fin(\cY)}\CCR^\SvN(\cY_1).\label{unio}\eeq
\eed

Clearly, $\CCR_\alg^\reg(\cY)$ is a $*$-algebra 
equipped with a $C^*$-norm.

\bed We define the {\em regular CCR $C^*$-algebra over $\cY$}
 as
\[\CCR^\reg(\cY):=
\left(\CCR_\alg^\reg(\cY)\right)^\cpl,\]
where the completion is with respect to the norm defined above.
\eed

Clearly, 
$\CCR^\reg(\cY)$ is a generalization of the Stone-von Neumann
algebra $\CCR(\cY)$ from
 Definition
\ref{stvnal}. 

We have an obvious extension of Proposition \ref{pres1}:

\bep 
Let $r\in
Sp(\cY)$. Then there exists a
unique  $*$-isomorphism
$\hat r:\CCR^\reg(\cY)\to\CCR^\reg(\cY)$ such that $\hat r(W(y))=W(ry)$,
$y\in \cY$, and if $\cY_1\in\Fin(\cY)$, then $\hat r$ restricted to
$\CCR^\SvN(\cY_1)\to\CCR^\SvN(r\cY_1)$ is  $\sigma$-weakly
continuous. 
 \label{pres2}\eep

Among the Bogoliubov automorphism one can distinguish
the {\em parity}, 
that is $\alpha:=\widehat{-\one}$. Clearly, $\alpha$ is an
involution.  Elements of 
$\CCR^{\reg}(\cY)$ fixed by $\alpha$ are called {\em even} and form a
subalgebra denoted 
$\CCR_0^\reg(\cY)$.

\bep Let
  $\cY\ni y\mapsto W^\pi(y)\in U(\cH)$ be a regular
CCR representation. Then there exists a unique 
 $*$-representation 
$\pi:\CCR^\reg(\cY)\to B(\cH)$ such that $\pi(W(y))=W^\pi(y)$,
$y\in\cY$, and which, for $\cY_1\in\Fin(\cY)$,
 is $\sigma$-weakly continuous on the subalgebras
$\CCR^\SvN(\cY_1)\subset\CCR^\reg(\cY)$.
Moreover, $\pi$ is isometric. \eep

\subsection{$C^*$-CAR algebras}

In the case of the CAR, there is an obvious well-known choice of a
$C^*$-algebra, which we recall in this subsection. It is discussed in
many places in the literature, see for instance \cite{Ar3,PR}.

Throughout the subsection we assume that $(\cY,\nu)$ is a euclidean
space. 
 
\bed \label{algo} The complex unital 
$*$-algebra generated by selfadjoint elements
 $\phi(y)$ depending linearly on $y\in\cY$ satisfying 
\beq[\phi(y_1),\phi(y_2)]_+= 2 y_1\nu y_2\one,\ \ \ y_1,y_2\in\cY
,\eeq
will be denoted by  $\CAR^\alg(\cY)$.\eed

It is easy to prove the following proposition:
\bep There exists a unique $C^*$-norm on 
 $\CAR^\alg(\cY)$. \eep

\bed We set
\[
\CAR^{C^*}(\cY):= \left(\CAR^\alg(\cY)\right)^{\cpl},
\]
where the completion is with respect to the $C^{*}$ norm defined
above. 
\eed

\ber 
In the literature, 
  $\CAR^{C^*}(\cY)$ is usually denoted
$\CAR(\cY)$.
Our more complicated notation is motivated by the fact that
there exist other 
natural  $*$algebras that describe the CAR over $\cY$. 
Another choice,
discussed in \cite{DG2,PR}, is the $W^*$-algebra 
 $\CAR^{W^*}(\cY)$, obtained by taking the weak closure in the GNS
representation for the unique tracial state on
 $\CAR^{C^*}(\cY)$. 
 \eer

Clearly, $\CAR^{C^*}(\cY)$  is a $C^{*}$ algebra.
It coincides with $\CAR^{C^*}(\cY^{\cpl})$. Hence it
is enough 
  to assume that $\cY$ is a real Hilbert space.

The following proposition describes
the so-called fermionic Bogoliubov automorphisms:

\bep If $r\in O(\cY)$,
there exists
a unique $*$-automorphism $\hat r$ of $\CAR^{C^*}(\cY)$ satisfying $\hat r(\phi(y))=\phi(ry)$.
\eep

Among the Bogoliubov automorphism a special role is played by the
parity, that is $\alpha:=\widehat{-\one}$. Clearly, $\alpha$ is an
involution.  Elements of 
$\CAR^{C^*}(\cY)$ fixed by $\alpha$ are called {\em  even} and form a
subalgebra denoted 
$\CAR_0^{C^*}(\cY)$.

The relationship between CAR representations  and the algebra 
 $\CAR^{C^*}(\cY)$ is given by the following proposition:

\bep If
\[\cY\ni y\mapsto \phi^\pi(y)\in B_\h(\cH)\]
is a CAR representation, then there exists a unique
$*$-homomorphism of $C^*$-algebras
\[
\pi:\CAR^{C^*}(\cY)\to B(\cH)
\]
 such that
$\pi(\phi(y))=\phi^\pi(y )$, $y\in\cY$.
\label{qwq1a}\eep

 $\CAR^{C^*}(\cY)$ belongs to one of the best known classes of
$C^*$-algebras, as seen from the following proposition:

\bep
If $\cY$ is infinite dimensional separable, then
  $\CAR^{C^*}
(\cY)$ is isomorphic
to the so-called
 uniformly hyperfinite $C^*$-algebra
 of the type $2^\infty$, sometimes
 denoted $UHF(2^\infty)$.\eep

\section{Quantization of 
neutral  systems}\label{c15b.1}\init

In this and the next section we
describe the positive quantization of   classical linear systems.
It is natural to divide the discussion into two sections: the first
 about neutral
and the second about charged systems. In both sections we describe
 bosonic
and fermionic systems. Then we consider discrete symmetries: the time
reversal 
and the charge reversal.

In the neutral formalism the classical phase
 space $\cY$ is real and 
is equipped with a
symplectic form $\omega$ in the bosonic case, resp.
with a positive scalar product
$\nu$
in the fermionic case. 
The dynamics describing the  {\em time evolution}
satisfies
 $r_{t}\in \Sp(\cY)$, resp. $r_{t}\in
O(\cY)$. The
problem adressed in this section is to 
find a CCR representation, resp. a CAR representation,
on a Hilbert
space $\cH$ and a positive  selfadjoint
operator $H$ on $\cH$ such that $\e^{\i tH}$ implements
$r_{t}$.

We will do it by finding a K\"{a}hler anti-involution that commutes
with the 
dynamics, 
and thus leads to a Fock representation in which the dynamics is
implementable.

It turns out that this is easy in the fermionic case. The bosonic
case is more technical. In particular,
one needs  to assume that the dynamics is {\em stable},
which roughly means that the classical Hamiltonian is positive.

One often assumes that the
dynamics $\{r_{t}\}_{t\in \rr}$ is a part of a larger group of symmetries $G$.
In other words, our starting point is  a homorphism of a  group $G$ into
 $\Sp(\cY)$, resp. $O(\cY)$. One often asks whether the action of $G$ can be
implemented in the Hilbert space $\cH$ by unitary or, sometimes, anti-unitary
operators.

\subsection{Neutral bosonic systems}

\subsubsection{Algebraic quantization of a symplectic dynamics}
\label{subs-sym-dyn}
Let $(\cY, \omega)$ be a symplectic space.
 Let $\rr\ni t\mapsto r_t\in
Sp(\cY)$  be
a 1-parameter  group. 

It is easy to describe the quantum counterpart of the above classical
dynamical system. We take one of the CCR algebras over 
 $(\cY, \omega)$,
 say $\CCR^\Weyl(\cY)$, and equip it with the group of
Bogoliubov automorphisms $\hat r_t$, defined by
\[\hat r_t(W(y))=W(r_ty),\ \ \ y\in\cY.\]

\subsubsection{Stable symplectic dynamics}
 
Typical symplectic dynamics  
that appear in physics have  {\em positive Hamiltonians}. We will call
such dynamics 
 {\em stable}.
We will see that (under some technical conditions) such dynamics lead to 
 uniquely defined Fock representations.

It is easy to make the concept of stability
precise if $\dim\cY<\infty$. 
In this case $\cY$ has a natural topology. We, of course, assume that
the dynamics $t\mapsto r_t$ is  continuous.
Let $a$ be its generator, so that
$r_t=\e^{ta}$. Clearly, the form $\beta$ defined by
\beq y_1\beta y_2:=y_1\omega
ay_2,\ \  y_1,y_2\in\cY,\eeq
is symmetric. We say that the group $t\mapsto r_t$ is {\em stable} if $\beta$
is strictly positive.

\subsubsection{K\"ahler structure for a 
weakly stable symplectic dynamics}
\label{c15b.1.5a}

To generalize the concept of stability to an infinite dimension, we need to
equip $(\cY,\omega)$ with a topology. There are various possibilities to do this, let
us consider the simplest one.

\bed
We say that $\left(\cY,\omega,\beta,(r_t)_{t\in\rr}\right)$
 is a {\em weakly stable
  dynamics}
if the following conditions are true:
\ben\item  $\beta$ is a positive definite symmetric form. We
equip $\cY$ with the norm $\|y\|_\en:=(y\cdot\beta y)^{\frac12}$. We
 denote  by
$\cY_\en$ the completion  of $\cY$ with respect to this norm.
\item  We assume
that   $t\mapsto r_t\in Sp(\cY)$ is bounded and strongly
continuous.
Thus we  can extend
$r_t$ 
to a strongly continuous group on $\cY_\en$ and define its generator $a$, so
that $r_t=\e^{ta}$. 
\item $\Ker a=\{0\}$, or equivalently, $\mathop{\bigcap}\limits_{t\in
\rr}\Ker(r_{t}-\one)=\{0\}$.
\item We assume that $\cY\subset \Dom\, a$ and
\beq\label{hammah}  y_1\beta y_2=y_1\omega
ay_2,\ \ y_1, y_2\in\cY.\eeq
\een
 If in addition $\omega$ is  bounded  for the
topology given by $\beta$, so that it can be extended to the whole $\cY_\en$,
we will say that the dynamics is {\em strongly stable}.
 In this case
$(\cY_{\en}, \omega)$ is a symplectic space.
\label{weakly stable}
\eed

Note that  $\beta$ has two roles: it endows $\cY$ with a topology and
it is the
Hamiltonian for $r_t$.

\bet Let  
 $\left(\cY,\omega,\beta,(r_t)_{t\in\rr}\right)$  be a weakly stable dynamics.
Then \ben \item $r_t$ are orthogonal transformations on the real Hilbert space
$\cY_\en$.
\item $a$ is anti-self-adjoint and $\Ker a=\{0\}$.
\item Let $h:=\sqrt{aa^*}=\sqrt{a^*a}$. The polar decomposition
 \[a=:h\ii=\ii h\]
defines an anti-involution $\ii$  on $\cY_\en$.
\item $h$ is a positive selfadjoint operator.
\item The dynamics is strongly stable iff $h\geq C$ for some $C>0$.\een
\label{coz}\eet

Recall that given a strictly 
positive operator $h$ on $\cY_\en$ we can define
a scale of Hilbert spaces $h^{s}\cY_\en$. Then $r_t$ and
$\ii$ are bounded on $\cY_\en\cap h^{s}\cY_\en$
for the norm of $h^{s}\cY_\en$. Let $r_{s,t}$ and $\ii_s$ denote
their extensions. Similarly, $a$ and $h$ are closable on
 $\cY_\en\cap h^{s}\cY_\en$ for the norm
 $h^{s}\cY_\en$. Let $a_s$, $h_s$ denote their
closures. Clearly, for any $s$, $a_s=h_s\ii_s=\ii_sh_s$ is the
polar decomposition, $\ii_s$ is an orthogonal anti-involution and
$r_{s,t}=\e^{t a_s}$ is an orthogonal 1-parameter group.

Let $\cdot_s$ denote the natural 
scalar product on  $h^{s}\cY_\en$. 
Let us express the scalar product and the symplectic form 
in terms of $(\cdot|\beta\cdot)$:
\begin{eqnarray*}
y_1\cdot_s y_2&=&y_1\beta h^{-2s}y_2=(h^{-2s}y_1)\beta y_2,\\
y_1\omega y_2&=&y_1\beta a^{-1}y_2=(a^{-1}y_1)\beta y_2
.
\end{eqnarray*}
Note that the symplectic form does not need to 
be everywhere defined,

Of particular interest for us is the case $s=\frac12$, for which we
introduce the notation
$\cY_\dyn:=h^{\frac12}\cY_\en$.
 In what follows we drop the subscript
$s=\frac12$ from $r_{s,t}$, $\ii_s$, $\cdot_s$, $a_s$ and $h_s$.

\bep $\cY_\dyn$ equipped with
 $(\cdot,\omega,\ii)$ is a 
 K\"ahler space. \eep


Clearly, $h$ is positive and we have a dynamics on $\cY_\dyn$
\[
 r_{t}= \e^{\ii t h}.
\]

\subsubsection{Fock quantization of symplectic
dynamics}\label{c15b.1.5}

 In this subsection we drop the subscript ${\scriptstyle\dyn}$
from $\cY_\dyn$.
  Let $\cZ$ be the  holomorphic subspace of $\cc\cY$
for the K\"{a}hler
anti-involution $\ii$ constructed in Thm. \ref{coz}. 

Clearly $h$ commutes with $\ii$, hence its complexification
$h_\cc$ preserves $\cZ$.
We set $h_\cZ:=h_\cc\Big|_\cZ$, which 
is a positive self-adjoint operator on $\cZ$
with $\Ker h_\cZ=\{0\}$.

Likewise $(r_t)_\cc$ preserves $\cZ$ and we have
\[(r_t)_\cc\Big|_\cZ=\e^{\i t h_\cZ}.\]

For $y\in\cY$ define the field operators
\[\phi(y)
:=a^*\left(\frac{\one-\i \ii}{2}y\right)+a\left(\frac{\one-\i
   \ii}{2}y\right) .\]
Then
 \beq\cY \ni y\mapsto 
 \e^{\i\phi(y)}\in U(\Gamma_\s(\cZ))\label{possi9}\eeq
 is a Fock CCR representation.
Introduce
 the positive operator $H:=\d\Gamma(h_\cZ)$ on $\Gamma_\s(\cZ)$.
We have 
\beq
\e^{\i tH}\phi(y)\e^{-\i tH}=\phi(r_{t} y).\label{possi1}
\eeq
\bed (\ref{possi9}) is called the {\em positive energy Fock
quantization}
for the weakly stable
dynamics $t\mapsto r_t$.\eed

\bex Let us describe a typical application of the neutral bosonic
formalism.

 Let
$\cY$ be the space of 
{\em real} solutions of the {\em  Klein-Gordon equation}
on the Minkowski space that have a compact support on any space-like hypersurface.  It is 
equipped with a natural symplectic form obtained by integrating the
well-known 
conserved current on any spacelike hypersurface.  $\cc\cY$
is the space of {\em complex} solutions and $\cZ$ is the space of
{\em positive frequency} solutions of the Klein-Gordon equation.

More generally, 
instead of the Minkowski space we can take a stationary globally
hyperbolic space-time and allow for a position-dependent mass.
\eex

\subsection{Neutral fermionic systems}

\subsubsection{Algebraic quantization of an orthogonal dynamics}
\label{subsec-ortho-dyn}

Let $(\cY,\nu)$ be a real Hilbert space. We think of it as the
phase space of a fermionic system. 
A strongly continuous  1-parameter 
group $\{r_{t}\}_{t\in \rr}$ with $r_{t}\in O(\cY)$ will be called an
{\em orthogonal dynamics}. We view it as a classical dynamical system.

We choose  $\CAR^{C^*}(\cY)$
as the field algebra of our system. It is
 equipped with the 1-parameter group of
Bogoliubov automorphisms $\hat r_t$, defined by
\[\hat r_t(\phi(y))=\phi({r_t}y),\ \ \ y\in\cY.\] 

In quantum physics only even fermionic operators are
observable. Therefore, it seems natural to use the 
even subalgebra   $\CAR_0^{C^*}(\cY)$ as the observable algebra.

\subsubsection{K\"ahler structure for
 a nondegenerate orthogonal dynamics}

Let $a$ be the generator of ${r_t}$, so that ${r_t}=\e^{ta}$ and
$a=-a^{\t}$.

\bed We say that the dynamics $t\mapsto r_t\in O(\cY)$ is
{\em nondegenerate} if
\beq
\Ker a=\{0\}, \hbox{ or equivalently }\bigcap_{t\in
\rr}\Ker(r_{t}-\one)=\{0\}.
\label{nondeg-bis}
\eeq\eed

\bet Set $h:=\sqrt{aa^*}=\sqrt{a^*a}$. The polar decomposition
 \[a=:h\ii=\ii h\]
defines an anti-involution $\ii$  on $\cY$,
 which is  K\"ahler for $\nu$.
\eet

Clearly, $h$ is positive and
\[
 r_{t}= \e^{\ii t h}.
\]

\subsubsection{Fock quantization of  orthogonal dynamics}\label{c15.2.1}

Let $\cZ$ be the holomorphic subspace of $\cc\cY$
for  the K\"{a}hler anti-involution
 $\ii$. 

The operator $h_\cc$ commutes with $\ii$. Hence, it
 preserves $\cZ$. We  set $h_\cZ:=
h_\cc\Big|_\cZ$. Note that $h_\cZ$ is positive.

Consider the Fock representation associated with the K\"ahler anti-involution
$\ii$
\beq\cY \ni y\mapsto \phi(y)
:=a^*\left(\frac{\one-\i \ii}{2}y\right)+a\left(\frac{\one-\i
  \ii}{2}y\right) 
\in B_\h(\Gamma_\a(\cZ)),\label{possi-bis}\eeq
and the positive operator $H:=\d\Gamma(h_\cZ)$ on $\Gamma_\a(\cZ)$.
We have 
\beq\e^{\i tH}\phi(y)\e^{\i tH}=\phi(r_{t}y).\label{possi1-bis}\eeq

\bed (\ref{possi-bis}) is called the {\em positive energy Fock
quantization} for the
dynamics $t\mapsto r_{t}$.\eed

\bex Let us describe a typical application of the neutral fermionic
formalism.

 Let
$\cY$ be the space of 
 solutions of the  {\em Dirac equation}
on the Minkowski space satisfying the {\em Majorana condition}
 that have a    compact support on any space-like hypersurface. 
 It is 
equipped with a natural scalar product obtained by integrating the
well-known conserved current on any spacelike hypersurface.  $\cc\cY$
is the space of all solutions, without imposing the Majorana
condition, 
 and $\cZ$ is the space of
{\em positive frequency} solutions of the Dirac equation.

More generally, 
instead of the Minkowski space we can take a stationary globally
hyperbolic space-time and allow for a position-dependent mass.
\eex

\subsection{Time reversal in neutral systems}

\subsubsection{Algebraic quantization of time reversal}
Let $(\cY,\omega)$ be a symplectic space
space with a dynamics $\rr\ni t\mapsto r_t\in
Sp(\cY)$ in the bosonic case, or 
let   $(\cY,\nu)$ be a real Hilbert space with a dynamics  
 $\rr\ni t\mapsto r_t\in O(\cY)$ in the fermionic case.

\bed A map $\tau\in L(\cY)$ 
 is called a {\em time
reversal} if   $\tau r_t=r_{-t}\tau$ and
\begin{eqnarray*}
\hbox{is anti-symplectic}&&\\
\hbox{and}\ \ \tau^2=\one& \hbox{in the bosonic case},&
\\\hbox{or}&&\\
\hbox{is orthogonal}&&\\
\hbox{and}\ \ \tau^2=\one\ \hbox{ or}\
\tau^2=-\one& \hbox{in the fermionic case}.&\label{qsx}
\end{eqnarray*}
\label{timerev}\eed

In the bosonic case we define an {\em antilinear}
 $*$-homomorphism of the algebra
$\CCR^\Weyl(\cY)$ by setting $\hat\tau(W(y)):=W(\tau y)$. Clearly,
$\hat\tau^2$ is the identity.

In the fermionic case, we 
we define an {\em antilinear} $*$-homomorphism of the algebra
$\CAR^{C^*}(\cY)$ by setting $\hat\tau(\phi(y)):=\phi(\tau
y)$. Clearly, restricted to $\CAR_0^{C^*}(\cY)$, 
$\hat\tau^2$ is the identity.

\subsubsection{Fock quantization of time reversal}

 In the bosonic case we assume that the
dynamics is weakly stable, in the fermionic case we assume that it is
nondegenerate. In both cases we can introduce $a$, $\ii$, $h$. Note that we have
\[
\tau a=-a\tau,\ \ \tau \ii=-\ii\tau,\ \ \tau h=h\tau.
\]
Recall that $\tau_{\bar\cc}$ denotes the {\em antilinear} extension of $\tau$
to $\cc\cY$. Note that $\tau_{\bar\cc}$ preserves $\cZ$. We write
$\tau_\cZ:=\tau_{\bar\cc}\Big|_{\cZ}$. Clearly, $\tau_\cZ$ is anti-unitary 
and 
\[ \tau_\cZ h_\cZ=h_\cZ\tau_\cZ .\]
\begin{eqnarray*}
\tau_\cZ^2=\one& \hbox{in the bosonic case},&\\
\tau_\cZ^2=\one\ \hbox{ or}\
\tau_\cZ^2=-\one& \hbox{in the fermionic case}.&
\end{eqnarray*}
Consider the positive energy quantization of the dynamics on the Fock
space $\Gamma_\sa(\cZ)$. On the quantum level the time reversal is
defined as the antiunitary  map $T:=\Gamma(\tau_\cZ)$.
We have
\[THT^{-1}=H,\ \ T\e^{\i tH}T^{-1}=\e^{-\i tH},\]
\[T\phi(y)T^{-1}=\phi(\tau y),\ \ y\in\cY.\]
Note that
\begin{eqnarray*}
T^2=\one& \hbox{in the bosonic case}&\\
T^2=\one\ \hbox{ or}\
T^2=I&
 \hbox{in the fermionic case,}&
\end{eqnarray*}
where $I$ is the parity operator defined in (\ref{c1para0}).

\section{Quantization  of charged systems}

In the charged formalism, the classical system is described
by a complex vector
space $\cY$. In  the bosonic case, it is
 equipped with an
anti-Hermitian form $(\cdot|\omega\cdot)$ -- we say that it is a {\em
  charged symplectic space}. 
The dynamics  $(r_t)_{t\in\rr}$
describing the  time evolution is assumed to 
preserve  $(\cdot|\omega\cdot)$, we say that 
$r_t$ is {\em charged symplectic}.
In the fermionic case it is equipped with 
 a positive scalar product
$(\cdot|\cdot)$ and without decreasing the generality we can assume
 that it is complete -- it is a complex Hilbert space.
The dynamics  $(r_t)_{t\in\rr}$
preserves
 $(\cdot|\cdot)$ -- it is unitary.

By a positive energy
quantization of a charged  classical system we mean finding 
 a charged CCR (resp. CAR) representation $y\mapsto \psi(y)$
 on a Hilbert space $\cH$ and a  positive selfadjoint
operator $H$ on $\cH$ such that $\e^{\i tH}$ implements
$r_{t}$.

The complex structure of $\cY$ is responsible for the action of a
$U(1)$ symmetry $(\e^{\i\theta})_{\theta\in[0,2\pi]}$. 
On the level of the Fock representation it is
implemented by 
 the charge operator $Q$.

Charged systems can be viewed as special cases of  neutral systems
equipped in addition
with a certain symmetry.
Recall that a homomorphism
 $U(1)\ni\theta\mapsto  u_\theta\in L(\cY)$
 is a {\em  $U(1)$ symmetry of charge 1}
 if there exists an anti-involution $\jq$ such that
 $u_\theta=\cos\theta\one+\sin\theta\ii_\ch$.
We assume that it preserves
 the symplectic, resp. euclidean form
$\omega$, resp. $\nu$, which is equivalent to saying that $\jq$ is
 pseudo-K\"ahler, resp. K\"ahler.
We also assume that   the dynamics $r_t$ commutes with the
symmetry, which is equivalent to saying that $\jq$ commutes with
$r_t$. 

If we  equip $\cY$ with a complex structure given by $\jq$,
then the symmetry
$u_\theta$ becomes just the multiplication by $\e^{\i \theta}$. 
It is then natural to replace the real
bilinear forms
$\omega$, resp. $\nu$  by closely related 
sesquilinear forms $(\cdot|\omega\cdot)$, resp.  $(\cdot|\cdot)$. The
invariance of the dynamics with respect to the charge symmetry is now
expressed by the fact that the dynamics is complex linear.

At the end of this section, we will discuss   the 
 {\em charge  reversal} and the {\em time reversal} for
charged systems.


\subsection{Charged bosonic systems}

\subsubsection{Algebraic quantization of a charged symplectic dynamics}

Let $(\cY,(\cdot|\omega\cdot))$  be a charged symplectic space.
Let $t\mapsto r_t\in ChSp(\cY)$ be a charged symplectic dynamics.

By taking $\Re(y_1|\omega y_2)$ we can view $\cY_\rr$ as a real
symplectic space. We choose
$\CCR^\reg(\cY_\rr)$ as 
the field algebra of 
our system. This algebra is generated (in the sense described 
in Subsect. \ref{s.reg})
by the Weyl elements denoted
$\e^{\i\psi(y)+\i\psi^*(y)}$, $y\in\cY$,  satisfying the relations
\[
\e^{\i\psi(y_1)+\i\psi^*(y_1)}\e^{\i\psi(y_2)+\i\psi^*(y_2)}
=\e^{-\i\Re(y_1|\omega y_2)}\e^{\i\psi(y_1+y_2)+\i\psi^*(y_1+y_2)}.\]

We can equip $\CCR^\reg(\cY_\rr)$ with the automorphism groups
$\widehat{\e^{\i\theta}}$ and $\hat r_t$ defined by
\begin{eqnarray*}
\widehat{\e^{\i\theta}}\left(\e^{\i\psi(y)+\i\psi^*(y)}\right)
&=&\e^{\i\psi(\e^{\i\theta}y)+\i\psi^*(\e^{\i\theta}y)},\\
\hat r_t\left(\e^{\i\psi(y)+\i\psi^*(y)}\right)
&=&\e^{\i\psi(r_ty)+\i\psi^*(r_ty)}.\end{eqnarray*}

As the  observable algebra it is natural to choose the so-called
 {\em gauge
  invariant regular CCR algebra}
 $\CCR_\gi^\reg(\cY)$, which is
defined as the set of elements of $\CCR^\reg(\cY_\rr)$ fixed by
$\widehat{\e^{\i\theta}}$. Note that
 $\CCR_\gi^\reg(\cY)$ is contained in the even algebra
 $\CCR_0^\reg(\cY_\rr)$  and
is preserved by the dynamics $\hat r_t$.

\ber In this subsection,
for the field algebra of our system we preferred to choose $\CCR^\reg(\cY_\rr)$
instead of $\CCR^\Weyl(\cY_\rr)$. This
is motivated by the fact that 
the only element left invariant by the gauge
symmetry $\widehat{\e^{\i\theta}}$ in
  $\CCR^\Weyl(\cY_\rr)$  is $\one$, whereas in the case 
of  $\CCR^\reg(\cY_\rr)$ we obtain a large gauge-invariant
algebra. \eer 

\subsubsection{Fock quantization of a charged symplectic dynamics}
\label{subsec-complex}

The concept of stability of dynamics in the charged case is analogous
to the neutral case.

\bed
We say that $\left(\cY,(\cdot|\omega\cdot),(\cdot|\beta\cdot),
(r_t)_{t\in\rr}\right)$
 is a {\em weakly stable
  dynamics}
if the following conditions are true:
\ben\item  $(\cdot|\beta\cdot)$
 is a positive definite sesquilinear form. We
equip $\cY$ with the norm $\|y\|_\en:=(y|\beta y)^{\frac12}$. We
 denote  by
$\cY_\en$ the  completion of $\cY$ with respect to this norm.
\item  We assume
that   $t\mapsto r_t$ is bounded and strongly
continuous.
Thus we  can extend
$r_t$ 
to a strongly continuous group on $\cY_\en$ and define its generator
$\i b$, so
that $r_t=\e^{t\i b}$. 
\item $\Ker b=\{0\}$, or equivalently, $\mathop{\bigcap}\limits_{t\in
\rr}\Ker(r_{t}-\one)=\{0\}$.
\item We assume that $\cY\subset \Dom b$ and
\beq\label{hammaha}  (y_1|\beta y_2):=\i(y_1|\omega
by_2),\ \ y_1, y_2\in\cY.\eeq
\een
 If in addition
\[|(y_1|\omega y_2)|\leq (y_1|\beta y_1)^{\frac12}
 (y_2|\beta y_2)^{\frac12},\]
so that $(\cdot|\omega\cdot)$
 can be extended to the whole $\cY_\en$,
we will say that the dynamics is {\em strongly stable}.
\label{strongly stable}
\eed

\bet Let  
 $\left(\cY,(\cdot|\omega\cdot),(\cdot|\beta\cdot)
,(r_t)_{t\in\rr}\right)$  be a weakly stable dynamics.
Then \ben \item $r_t$ are unitary transformations on the 
 Hilbert space
$\cY_\en$.
\item $b$ is  self-adjoint and $\Ker b=\{0\}$.
\een
\label{coz1}\eet

Set $q:=\sgn b$,
 $\ii:=\i\,\sgn b$ and
 $h:=|b|$. Clearly $h$ is positive,  and $r_t=\e^{t\ii h}$.

Set
$\cY_\dyn:=h^{\frac12}\cY_\en$. 
As in Subsect. \ref{c15b.1.5a}, 
we can view $r_t$, $\ii$, $b$ and $h$ as 
defined on $\cY_\dyn$. In what follows we drop the subscript
${\scriptstyle \dyn}$
from $\cY_\dyn$.

 Let $\one_\pm:=
\one_{]0,\infty[}(\pm b)=\one_{\{\pm1\}}(q)$,
 $\cY_{\pm}:=\Ran\one_\pm$.
Let $\cZ$ denote the space $\cY$ equipped with the complex structure
given by $\ii$. (In other words,
$\cZ:=\cY_{+}\oplus\bar\cY_{-}$).

The operators $h$, $q$ and $b$ preserve
$\cY_\pm$. Hence they can be viewed as
 complex linear operators
on $\cZ$ as well, in which case they will be denoted
 $h_\cZ$, $q_\cZ$ and $b_\cZ$.

Consider
the space $\Gamma_\s(\cZ)$.
For $y\in\cY$, let
 us introduce the {\em charged fields} on $\cY$,
 which are closed operators on $\Gamma_\s(\cZ)$ defined by
\begin{eqnarray}
\psi^*(y)&=&a^*\left(\one_{+}y\right)+a\left(\bar{\one_{-}y}\right),
\nonumber\\
\psi(y)
&=&a\left(\one_{+}y\right)+a^*\left(\bar{\one_{-}y}\right).
\label{fockq1}
\end{eqnarray}
We obtain a charged CCR representation
\beq\cY\ni y\mapsto\psi(y)\in Cl(\Gamma_\s(\cZ)).
\label{chacha2}\eeq

Define the self-adjoint operators on  $\Gamma_\s(\cZ)$:
\[H:=\d\Gamma(h_\cZ),\ \ Q:=\d\Gamma(q_\cZ).\]
 Clearly,
\[
\e^{\i tH}\psi(y)\e^{-\i tH}= \psi(\e^{\i tb}y), \ \ \e^{\i
\theta Q}\psi(y)\e^{-\i \theta Q}= \psi(\e^{\i\theta}y),\ \ y\in\cY.
\]
\bed (\ref{fockq1}) is called the {\em positive energy Fock
quantization} for the
dynamics $t\mapsto r_{t}$.\eed

\bex Let us describe a typical application of the charged bosonic
formalism.

 Let
$\cY$ be the space of 
{\em complex} solutions of the  {\em Klein-Gordon equation}
that have a compact support on any space-like hypersurface.
We obtain $\cZ$
by switching the sign of 
the imaginary unit on {\em negative frequency} 
  solutions. During  quantization, this means that for negative
frequency solutions we switch the role of creation and annihilation
operators.

As in the neutral case
we can  allow for  a stationary globally
hyperbolic space-time and a position-dependent mass. In
addition, we can include a time-independent
 external vector potential.
\eex

\subsection{Charged fermionic systems}
\subsubsection{Algebraic quantization of a unitary dynamics}

Let $\left(\cY, (\cdot|\cdot)\right)$  be a complex Hilbert space
describing a charged fermionic system.
A strongly continuous  1-parameter 
group $\{r_{t}\}_{t\in \rr}$ with $r_{t}\in U(\cY)$ will be called a
{\em unitary dynamics}.

Clearly, by taking the real scalar product $y_1\nu
y_2:=\Re(y_1|y_2)$ we can view $\cY_\rr$
 as a real Hilbert space.
We can associate to our system the field algebra
$\CAR^{C^*}(\cY_\rr)$
 with distinguished elements $\psi(y)$. We can equip
it  with the automorphism group
$\widehat{\e^{\i\theta}}$ and $\hat r_t$ defined by
\begin{eqnarray*}
\widehat{\e^{\i\theta}}(\psi(y))&=&\psi(\e^{\i\theta}y),\\
\hat r_t(\psi(y))&=&\psi(r_t y).\end{eqnarray*}

Similarly as in the bosonic case, for 
the  observable algebra we choose the so-called
{\em gauge-invariant CAR algebra} $\CAR_\gi^{C^*}(\cY)$, which
 is
defined as the set of elements of
$\CAR^{C^*}(\cY_\rr)$ fixed by
$\widehat{\e^{\i\theta}}$.
Note that $\CAR_\gi^{C^*}(\cY)$
is contained in the even algebra
$\CAR_0^{C^*}(\cY_\rr)$ and is preserved by the
 dynamics $\hat r_t$.

\subsubsection{Fock quantization of a unitary dynamics}
\label{complex-fermi}

Let $\i b$ be the generator of ${r_t}$, so that ${r_t}=\e^{\i tb}$ and 
$b$ is self-adjoint.

\bed We say that the dynamics $t\mapsto r_t\in U(\cY)$ is
{\em nondegenerate} if
\beq
\Ker b=\{0\}, \hbox{ or equivalently }\bigcap_{t\in
\rr}\Ker(r_{t}-\one)=\{0\}.
\label{nondeg-bis1}
\eeq\eed

Set $q:=\sgn b$,
 $\ii:=\i\,\sgn b$ and
 $h:=|b|$. Clearly $h$ is positive,  and $r_t=\e^{t\ii h}$.
 Let $\one_\pm:=
\one_{]0,\infty[}(\pm b)=\one_{\{\pm1\}}(q)$,
 $\cY_\pm:=\Ran\one_\pm$.
Let $\cZ$ denote the space $\cY$ equipped with the complex structure
given by $\ii$. (In other words,
$\cZ:=\cY_+\oplus\bar\cY_-$).

The operators $h$, $q$ and $b$ preserve
$\cY_\pm$. Hence they 
can be also viewed as complex linear operators
on $\cZ$ as well, in which case they will be denoted
 $h_\cZ$, $q_\cZ$ and $b_\cZ$.

Consider
the space $\Gamma_\a(\cZ)$.
For $y\in\cY$, let
 us introduce the {\em charged fields} on $\cY$,
 which are closed operators on $\Gamma_\a(\cZ)$
\begin{eqnarray}
\psi^*(y)&=&a^*\left(\one_{+}y\right)+a\left(\bar{\one_{-}y}\right),\\
\psi(y)
&=&a\left(\one_{+}y\right)+a^*\left(\bar{\one_{-}y}\right).
\label{fockq2}
\end{eqnarray}
We obtain a charged CAR representation
\beq\cY\ni y\mapsto\psi(y)\in B(\Gamma_\a(\cZ)).
\label{chacha2a}\eeq

Define the self-adjoint operators on  $\Gamma_\a(\cZ)$
\[H:=\d\Gamma(h_\cZ),\ \ Q:=\d\Gamma(q_\cZ).\]
 Clearly,
\[
\e^{\i tH}\psi(y)\e^{-\i tH}= \psi(\e^{\i tb}y), \ \ \e^{\i
\theta Q}\psi(y)\e^{-\i \theta Q}= \psi(\e^{\i\theta}y),\ \ y\in\cY.
\]

\bed (\ref{fockq2}) is called the {\em positive energy Fock
quantization} for the
dynamics $t\mapsto r_{t}$.\eed

\bex Let us describe a typical application of the charged fermionic
formalism.

 Let
$\cY$ be the space of 
 solutions of the {\em  Dirac equation}
that have a compact support on any space-like hypersurface.
 We obtain $\cZ$
by switching the sign of the imaginary unit on 
{\em negative frequency}
  solutions. During quantization, this means that for negative
frequency solutions we switch the role of creation and annihilation
operators.

As in the neutral case
we can  allow for  a stationary globally
hyperbolic space-time and a position-dependent mass. In
addition, we can include
 a time-independent external vector potential.
\eex

\subsection{Charge reversal}\label{c15.4.2}

\subsubsection{Algebraic quantization of charge reversal}
 Let $(\cY, (\cdot|\omega\cdot ))$
 be a charged symplectic space in the bosonic case, or let
$(\cY, (\cdot|\cdot))$ be a 
complex Hilbert  space in the fermionic case.

\bed
 We say that $\chi$ is a charge reversal iff $\chi$ is
 anti-linear, $\chi^2= \one$ or
$\chi^2=-\one$, and
\begin{eqnarray*}\nonumber
(\chi y_1|\omega\chi y_2)&=&\bar{( y_1|\omega y_2)},
\ \ \hbox{($\chi$ is anti-charged symplectic) in the bosonic case};\\
(\chi y_1|\chi y_2)&=&\bar{( y_1| y_2)},\ \ \ \ \hbox{($\chi$ is
anti-unitary) in the fermionic case}.
\end{eqnarray*}
\eed

\bed Suppose that $\{r_{t}\}_{t\in \rr}$ is a 
charged symplectic (resp. unitary) dynamics.
  We say that the dynamics is {\em invariant under
the charge reversal $\chi$} if 
\[
\chi
r_{t}= r_{t}\chi, \ \ t\in \rr.
\] 
Similarly, if we have a group of symmetries  $\{r_g\}_{g\in G}$
we say that it is invariant under charge reversal $\chi$ iff
 $r_g\chi=\chi r_g$, $g\in G$. \eed

In the bosonic case we define the (linear) automorphism $\hat \chi$
 of the algebra
$\CCR^\reg(\cY_\rr)$ by
\begin{eqnarray*}
\hat \chi\left(\e^{\i\psi(y)+\i\psi^*(y)}\right)
&=&\e^{\i\psi(\chi y)+\i\psi^*(\chi y)}.\end{eqnarray*}
It restricts to an automorphism of
$\CCR_\gi^\reg(\cY)$.

In the fermionic case we define the (linear) automorphism $\hat\chi$
of the algebra $\CAR^{C^*}(\cY_\rr)$ 
by $\hat\chi(\psi^*(y))=\psi(\chi y)$.
It restricts to an automorphism of
$\CAR_\gi^{C^*}(\cY)$.

\subsubsection{Fock quantization of charge reversal}

 In the  bosonic case, assume
that the dynamics is weakly stable.
 In the fermionic case assume it is
nondegenerate. Let $b$, $h$, $q$ etc. be constructed as before.
In both bosonic and fermionic cases, it follows that
\begin{eqnarray*}
 \chi h=h\chi,&
\chi b=-b\chi,&\chi q=-q\chi,\ \ \chi\ii=\ii\chi.\end{eqnarray*}
We denote $\chi_\cZ$ the map $\chi$ considered on $\cZ$.
 Note that $\chi_\cZ$, unlike $\chi$, is unitary.
We second-quantize it by the unitary $C:=\Gamma(\chi_\cZ)$.
We have
\[CHC^{-1}=H,
\ \ CQC^{-1}=-Q,\]
\[C\psi^*(y)C^{-1}=\psi(\chi y).\]
Note that
\begin{eqnarray*}
C^2=\one& \hbox{ or}&
C^2=I
.\end{eqnarray*}

\subsubsection{Neutral subspace}

Assume that $\chi^2=\one$. We
can then  define the space $\cY_\chi:=\{y\in\cY\ :\ 
y=\chi y\}$ and restrict the dynamics
and the symmetry group to
 $\cY_\chi$. One can call $\cY_\chi$ the {\em neutral subspace} of
$\cY$.  (In the fermionic case it is also called the {\em Majorana 
subspace}). Note that $\cY=\cY_\chi\oplus\i\cY_\chi$,
hence the system can be viewed as a couple of neutral systems.

Let us describe the converse
 construction. Suppose  that we have  a  neutral system  
 $(\cY, \omega)$ or $(\cY, \nu)$  equipped with the dynamics $t\mapsto r_t$. We can extend it
to a charged system as follows. We consider the complexified space
$\cc\cY$ equipped with the natural conjugation denoted by the ``bar''. We
equip it 
 with the anti-hermitian form, resp. scalar product
\begin{eqnarray*}
(y_1|\omega y_2)&:=&\bar y_1\omega y_2,\\
\ \ \ \hbox{or}\ \ 
(y_1|y_2)&:=&\bar y_1\nu y_2,\ \ y_1,y_2\in\cc\cY.
\end{eqnarray*}
We extend the dynamics $r_t$ to  $(r_t)_\cc$ on
$\cc\cY$. Clearly, $(r_t)_\cc$ is a charged symplectic, resp. unitary
dynamics  and
the complex conjugation $\chi y:=\bar y$ is a charge
reversal satisfying $\chi^2=\one$. One gets back the original
system by the restriction to the neutral subspace.

\subsection{Time reversal in charged systems}

\subsubsection{Algebraic quantization of time reversal}

 Let $(\cY, (\cdot|\omega\cdot ))$
 be a charged symplectic space  in the bosonic case, or let
$(\cY, (\cdot|\cdot))$ be a 
complex Hilbert  space in the fermionic case.
\bed
 We say that $\tau\in L(\cY)$ is a time reversal iff $\tau
 r_t=r_{-t}\tau$,  $\tau$ is
 anti-linear, $\tau^2 = \one$  or
$\tau^2=-\one$, and
\begin{eqnarray*}\nonumber
(\tau y_1|\omega\tau y_2)&=&-\bar{( y_1|\omega y_2)},\ \ \hbox{($\tau$
  is anti-charged anti-symplectic) in the bosonic case};\\ 
(\tau y_1|\tau y_2)&=&\bar{( y_1| y_2)},\ \ 
\hbox{($\tau$ is anti-unitary) in the fermionic case}.
\end{eqnarray*}
\eed

In the bosonic case
 we define the {\em anti-linear} $*$-automorphism $\hat \tau$
 of the algebra
$\CCR^\reg(\cY_\rr)$ by
\begin{eqnarray*}
\hat \tau\left(\e^{\i\psi(y)+\i\psi^*(y)}\right)
&=&\e^{-\i\psi(\tau y)-\i\psi^*(\tau y)}.\end{eqnarray*}
It restricts to an anti-linear $*$-automorphism of
$\CCR_\gi^\reg(\cY)$.

In the fermionic case we define the {\em anti-linear} $*$-automorphism
$\hat\tau$ 
of the algebra $\CAR^{C^*}(\cY_\rr)$ 
by $\hat\tau(\psi(y))=\psi(\tau y)$.
It restricts to an automorphism of
$\CAR_\gi^{C^*}(\cY)$.

\subsubsection{Fock quantization of time reversal}

Clearly, we have $ \tau q=q\tau$.
Thus $\tau\cY_+=\cY_+$, $\tau\cY_-=\cY_-$.

Let $\tau_\cZ$ denote $\tau$ considered on $\cZ$.
It is anti-linear. We second-quantize
$\tau$ by the anti-unitary $T:=\Gamma(\tau)$. We obtain
\begin{eqnarray*}
THT^{-1}=H,&& T\e^{\i tH}T^{-1}=\e^{-\i tH},\\
TQT^{-1}=Q,&& T\e^{\i\theta Q}T^{-1}=\e^{-\i\theta Q}.\\
T\psi(y)T^{-1}=\psi(\tau y),&&T\psi^*(y)T^{-1}=\psi^*(\tau y)
.\end{eqnarray*}
\begin{eqnarray*}
T^2=\one& \hbox{ or}&
T^2=I
.\end{eqnarray*}

\subsubsection{Commutation between charge and time reversal}

It is natural to assume that  on the observable
algebra 
$(\hat\chi\hat\tau)^2$ is the identity. This is guaranteed if $(\chi\tau)^2$
equals $\one$ or $-\one$.  This leads
to the following \hbox{(anti-)commutation} relations for $\chi$ and $\tau$:
\begin{eqnarray*}\nonumber
\tau\chi\ =\ \chi\tau\ \hbox{ or}\
\tau\chi&=&-\chi\tau.
\end{eqnarray*}
However, we are free to multiply either $\chi$ or $\tau$ by $\i$.
Therefore,
 possibly after a redefinition of $\chi$ or $\tau$,
 we can always assume that 
\beq \tau\chi=\chi\tau.\eeq
Thus we have 3 commuting symmetries: $\chi$, $\tau$ and
$\chi\tau$. They satisfy  one of the following sets of relations:
\begin{eqnarray*}
\chi^2=\one,&\tau^2=\one,&(\chi\tau)^2=\one;\\
\chi^2=-\one,&\tau^2=-\one,&(\chi\tau)^2=\one;\\
\chi^2=\one,&\tau^2=-\one,&(\chi\tau)^2=-\one;\\
\chi^2=-\one,&\tau^2=\one,&(\chi\tau)^2=-\one.
\end{eqnarray*}


\end{document}